\documentclass{aa}  
\usepackage{graphicx}
\usepackage{txfonts}
\usepackage{hyperref}
\newcommand{\rf}[1]{{#1}}

\newcommand{\nsamp}{12}  % number of FRB sightlines used here

\newcommand{\pccc}{pc\,cm$^{-3}$}

\newcommand{\dm}{\ensuremath{\mathrm{DM}}}

\newcommand{\dmfrb}{\ensuremath{\mathrm{DM}_\mathrm{FRB}}}

\newcommand{\dmcosmic}{\ensuremath{\mathrm{DM}_\mathrm{cosmic}}}
\newcommand{\dmacosmic}{\ensuremath{\langle\dmcosmic\rangle}}
\newcommand{\dmcosmicmin}{\ensuremath{\mathrm{DM}_\mathrm{cosmic}^\mathrm{min}}}

\newcommand{\dmmw}{\ensuremath{\mathrm{DM}_\mathrm{MW}}}
\newcommand{\dmmwmin}{\ensuremath{\mathrm{DM}_\mathrm{MW}^\mathrm{min}}}
\newcommand{\dmmwism}{\ensuremath{\mathrm{DM}_\mathrm{MW}^\mathrm{ISM}}}
\newcommand{\dmmwhalo}{\ensuremath{\mathrm{DM}_\mathrm{MW}^\mathrm{halo}}}
\newcommand{\dmunits}{\ensuremath{{\rm pc \, cm^{-3}}}}

\newcommand{\dmhobs}{\ensuremath{\mathrm{DM}_\mathrm{host}^\mathrm{obs}}}  % host DM in observer frame
\newcommand{\dmhmacq}{\ensuremath{\mathrm{DM}_\mathrm{host}^\mathrm{Macquart}}} % host DM in rest frame
\newcommand{\dmhmax}{\ensuremath{\mathrm{DM}_\mathrm{host}^\mathrm{max}}} % host DM in rest frame
\newcommand{\dmhrest}{\ensuremath{\mathrm{DM}_\mathrm{host}^\mathrm{rest}}} % host DM in rest frame
\newcommand{\dmhfix}{\ensuremath{\mathrm{DM}_\mathrm{host,fix}^\mathrm{rest}}}  % fixed host DM value
\newcommand{\dmhost}{\ensuremath{\mathrm{DM}_\mathrm{host}}}
\newcommand{\dmhdirect}{\ensuremath{\mathrm{DM}_\mathrm{host}^\mathrm{direct}}} % host DM in rest frame
\newcommand{\dmhism}{\ensuremath{\mathrm{DM}_\mathrm{host}^\mathrm{ISM}}} % host DM in rest frame
\newcommand{\dmhilocal}{\ensuremath{\mathrm{DM}_\mathrm{host}^\mathrm{ISM,local}}} % host DM ISM from local Halpha
\newcommand{\dmhiglobal}{\ensuremath{\mathrm{DM}_\mathrm{host}^\mathrm{ISM,global}}} % host DM from global halpha
\newcommand{\dmhhalo}{\ensuremath{\mathrm{DM}_\mathrm{host}^\mathrm{halo}}} % host DM in rest frame
\newcommand{\halpha}{\ensuremath{\mathrm{H}\alpha}}
\newcommand{\hbeta}{\ensuremath{\mathrm{H}\beta}}
\newcommand{\hgamma}{\ensuremath{\mathrm{H}\gamma}}

\newcommand{\sbha}{\ensuremath{S(\halpha)}} % Surface brightness of Halpha
\newcommand{\dha}{19} % Surface brightness of Halpha
\newcommand{\ddm}{14}

\newcommand{\avgdirect}{80 \pm 11}  % average DM_direct +/- its error
\newcommand{\stddirect}{38}  %37.7
\begin{document}

   \title{Empirical estimation of host galaxy dispersion measure towards well localized fast radio bursts}

   \author{Lucas Bernales--Cortes\inst{1}
   \and 
   Nicolas Tejos\inst{1}
   \and
   J. Xavier Prochaska\inst{2,3,4,5}
   \and
   Ilya S. Khrykin\inst{1}
   \and
   Lachlan Marnoch\inst{6,7,8}
   \and\\
   Stuart~D.~Ryder\inst{6,7}
   \and
   Ryan~M.~Shannon\inst{9}
   }
   \institute{
   Instituto de F\'isica, Pontificia Universidad Cat\'olica de Valpara\'iso, Casilla 4059, Valpara\'iso, Chile\\ \email{lucas.bernales.c@mail.pucv.cl; nicolas.tejos@pucv.cl}
   \and
    University of California, Santa Cruz, 1156 High St., Santa Cruz, CA 95064, USA
    \and
    Kavli IPMU (WPI), UTIAS, The University of Tokyo, Kashiwa, Chiba 277-8583, Japan
    \and
    Division of Science, National Astronomical Observatory of Japan, 2-21-1 Osawa, Mitaka, Tokyo 181-8588, Japan
    \and
    Simons Pivot Fellow
    \and
   School of Mathematical and Physical Sciences, Macquarie University, NSW 2109, Australia
   \and
   Astrophysics and Space Technologies Research Centre, Macquarie University, Sydney, NSW 2109, Australia
   \and
   Australia Telescope National Facility, CSIRO Space \& Astronomy, Box 76 Epping, NSW 1710, Australia
   \and
   Centre for Astrophysics and Supercomputing, Swinburne University of Technology, Hawthorn, VIC, 3122, Australia
   }

   \date{\today}

% \abstract{}{}{}{}{} 
% 5 {} token are mandatory

  \abstract
   {\rf{Fast radio bursts} (FRBs) are very energetic pulses in the radio wavelengths that  have an unknown physical origin. These can be used to study the intergalactic medium (IGM) thanks to their dispersion  measure (DM). The DM has several contributions that can be measured (or estimated), including the contribution from the host galaxy itself, \dmhost. This \dmhost\ is generally difficult to measure, thus limiting the use of FRBs as cosmological probes and/or for understanding their physical origin(s). }
   {In this work, we empirically estimate \dmhost\ for a sample of \rf{twelve} galaxy hosts of well-localized FRB at $0.11<z<0.53$, using a direct method based solely on the properties of the host galaxies themselves, referred to as \dmhdirect. We also \rf{explore possible} correlations between \dmhost\ and some key global properties of galaxies.}
   {We use VLT/MUSE observations of the FRB hosts for estimating our empirical \dmhdirect. The method relies on estimating the DM contribution of both the FRB host galaxy's interstellar medium (\dmhism) and its halo (\dmhhalo) separately. 
   For comparison purposes, we also provide an alternative indirect method to estimate \dmhost\ based on \rf{the Macquart relation} (\dmhmacq).} 
   {We find an average $\langle \dmhost\rangle = \avgdirect$\,\pccc\ \rf{with a standard deviation of \stddirect\,\pccc\ }(in the rest-frame) based on our direct method, with a systematic uncertainty of $\sim 30$\%. This is larger than the typically used value of $50$\,\pccc\ but consistent within uncertainties. We report positive correlations between \dmhost\ and both the stellar masses and the star-formation rates of their hosts galaxies. In contrast, we do not find any strong correlation between \dmhost\ and neither redshift nor the projected distances to the center of the FRB hosts. Finally, we do not find any strong correlation between \dmhdirect\ and \dmhmacq, although the average values of the two are consistent within uncertainties.}
  % conclusions heading (optional), leave it empty if necessary.
   {Our reported correlations between \dmhdirect\ and stellar masses and/or the SFRs of the galaxies could be used in future studies to improve the priors used in establishing \dmhost\ for individual \rf{FRBs}. Similarly, such correlations and the lack of a strong redshift evolution can be used to constrain models for the progenitor of FRBs, e.g. by comparing them with theoretical models. However, the lack of correlation \dmhdirect\ and \dmhmacq\ \rf{indicates} that there may still be contributions to the DM of FRBs not included in our \dmhdirect\ modeling, e.g. large DMs from the immediate environment of the FRB progenitor and/or intervening large-scale structures not accounted for in \dmhmacq.}


   \maketitle
%
%-------------------------------------------------------------------

\section{Introduction}

    A \rf{fast radio burst} (FRB) is a \rf{short (of the order of milliseconds), very energetic event ($\gtrsim 10^{39}$\,erg) observed} at radio wavelengths \citep{LorimerBurst, PetroffReview2019,cordes2019,Petroff2022}. The progenitor and emission processes of FRBs remain unknown, although models implicating young magnetars \citep{magnetar,lensmagnetar}, or supernovae \citep{QNova} have been favored over the last years (see \cite{platts19} for a compendium of theories).  One of the key observational properties of FRBs is the dispersion measure (\dm), which provides an estimate \rf{of the integrated column density} of ionized matter along the line-of-sight to the pulse progenitor. The \dm\ is defined as:
    
    \begin{equation}
        \text{DM} = \int \frac{n_e(l)}{1+z} \text{d} l \ \text{,}
    \end{equation}

\noindent where $n_e$ is the density of free electrons, $l$ is the line-of-sight path, and $z$ is the redshift. 
    
Having well-localized ($\lesssim1''$) FRBs with identified host  galaxies allows us to study and understand the observed \dmfrb\ of a given FRB in terms of the contributions from different cosmological scales \citep{CosmicDM}. We can expand \dmfrb\ as

    \begin{equation}
        \dmfrb =
          \dmmw + \dmcosmic(z)
          + \frac{\dmhrest}{1+z}
        \ \text{,}
        \label{eq:dm_obs}
    \end{equation}

\noindent with

\begin{equation}
\dmmw \equiv \dmmwism + \dmmwhalo \;\;,
\label{eq:mw}
\end{equation}

\noindent where \dmmwism\ is the contribution from the Galactic interstellar medium (ISM), \dmmwhalo\ is the contribution from the Milky Way halo, 
\dmhrest\ 
is the contribution from the host galaxy in its rest-frame
including its halo, ISM and any gas local to the event itself, and 
\dmcosmic$(z)$\  
is the contribution from all other extragalactic gas, like the ionized gas in the large scale structure of the Universe and intervening halos of galaxies intersecting the FRB sightline (if any).

The ISM of the Milky Way is typically characterized using observations of the pulsar population, leading to \rf{the NE2001 \citep{NE2001-II} and YMW16 \citep{YMW16} Galactic electron density models.}
The average \dmcosmic\ component, 
\dmacosmic, follows the Macquart relation \citep{CosmicDM}, which relates the redshift of the FRB host galaxy with the DM$_{\rm cosmic}(z)$.

For the \dmhrest\ contribution, it is common to use a fixed value (e.g. $\dmhfix =50 \, \dmunits$\rf{; }\citealp{Arcus2020}, and references therein), or estimate it from a probability density function (PDF) expectation in the context of a Bayesian framework (typically log-normal, e.g. \citealp{CosmicDM, james+23};\citealp{james+2021,khrykin2024b}). Alternatively, other authors \citep[e.g.][]{HostR3,niu2022,lee+23} estimate it by re-writing Equation~\ref{eq:dm_obs} to solve for \dmhrest\ (see Section~\ref{sec:macquart}).
However, a single \dmhrest\ value, or range (e.g. \citealp{james+2021}), for {\it all} FRB hosts (even at the same redshift) is not physically motivated because one expects local factors to directly affect the \dmhrest\ value such as local density, clumpiness, ionization state, etc.; similarly, the mass of the galaxy, the temperature of its halo, and the 3D position of the FRB source with respect to the observer must also affect the intrinsic \dmhost\ value.  

More direct estimations of \dmhost\footnote{In the following, we will refer to DM$_{\rm host}^{\rm rest}$ as \dmhost\ unless otherwise noted. The relation between \dmhost\ in the observer-frame versus in the source-frame is ${\rm DM_{\rm host}^{\rm obs}=DM_{host}^{\rm rest}}/(1+z)$.}  exist, which explicitly attempt to take into account the contribution of the galaxy host ISM via nebular emission lines. For instance \cite{Tendulkar2017} obtained an estimate of $55 \leq \dmhobs \leq 225~\dmunits$ at $z=0.19$ based on \halpha\ for FRB20121102A. Similarly, \cite{ChittidiFRB190608} obtained  $\dmhobs= 150\pm 45$~\pccc\ with the same method for the galaxy host of FRB20190608B at $z=0.117$. \cite{FRB180924} estimate $30 \leq \dmhobs \leq 81~\dmunits$ for the host galaxy of FRB20180924B at $z=0.32$ using a Bayesian framework. In all these methods the uncertainty on \dmhost\ is large, and its value is difficult to determine precisely. It is therefore important to investigate whether we can better (if at all) determine \dmhost, which motivates our present study.

Our goal is to obtain an empirical estimate \rf{for this} \dmhost\ component, referred to as \dmhdirect, in a sample of several FRBs and analyzed in a homogeneous manner.
In principle, in order to fully understand how the host galaxy affects \dmhost, it is essential to consider not only the global properties of the galaxy but also the \rf{local environment of the FRB}. Indeed, the specific region from which the FRB originates can significantly influence the interpretation of observational data and the understanding of the physical mechanisms involved \citep{mannings21,Woodland24}.
For addressing this, in this work we explore if there is any difference in using nebular gas emission from the ``global'' ISM (full galaxy) or ``local'' to the \rf{source} (at the position of the FRB). Furthermore, we also investigate some possible relations between this \dmhdirect\ and internal properties of the host galaxies like stellar mass, star formation rate (SFR), as well as the geometry between the FRB and the host galaxy (e.g. projected distance of the FRB position and the galaxy center). 

Our paper is structured as follows.
In Section~\ref{sec:data} we present the observational data utilized and the sample analyzed in this study, including some key properties of the host galaxies. In Section~\ref{sec:analysis} we elaborate on our direct method as well as a previous (indirect) method to estimate the \dmhost.
In Section~\ref{sec:discussion} we report the results and discuss them, while in Section~\ref{sec:summary} we provide the summary and main conclusions. In this work, we assume a flat $\Lambda$CDM cosmology with parameters consistent with the latest Planck 2018 measurements \citep{planck18}.

%--------------------------------------------------------------------
\section{Data}
\label{sec:data}

\subsection{Sample} 
\label{sec:sample}

We use a sample of \rf{eleven} well-localized FRBs detected by the Commensal Real-Time ASKAP Fast-Transients (CRAFT; \citealp{CRAFT}) survey with identified host galaxies for which we have VLT/MUSE data (see below). We require better than 1\arcsec\ of precision in the FRB localization, and to make a confident host galaxy association we require a Probabilistic Association of Transients to their Hosts (PATH; \citealp{PATH}) posterior probability greater than 90$\%$. In addition to ASKAP FRBs, we have also included FRB20210410D detected by the More TRAnsients and Pulsars (MeerTRAP; \citealp{MeerTRAP}) project which also satisfies the above criteria. Our final sample is composed of \rf{twelve} FRB host galaxies and is summarized in Table~\ref{tab:sample}. 

\begin{table*}[ht]
\centering
\begin{tabular}{ccccccc}
\hline
\hline
    FRB name &  Redshift &  DM$_{total}$ &   RA FRB &  Dec FRB & Host probability &                Reference \\
& & (pc cm$^{-3}$)& (deg) & (deg) & ($\%$)  \\ (1) & (2) & (3) & (4) & (5) & (6) & (7) \\
\hline
FRB20180924B &   0.3212 &       362 & 326.1052 & -40.9000 &            99.9  &         \cite{FRB180924} \\
FRB20190102C &   0.2912 &       365 & 322.4157 & -79.4757 &            98.0  &      \cite{Bhandari2020} \\
FRB20190608B &   0.1178 &       340 & 334.0199 &  -7.8982 &             99.9 & \cite{ChittidiFRB190608} \\
FRB20190611B &   0.3778 &       333 & 320.7456 & -79.3976 &            100.0 &           \cite{Day2020} \\
FRB20190711A &   0.5217 &       595 & 329.4192 & -80.3580 &            97.3  &        \cite{Heintz2020} \\
FRB20190714A &   0.2365 &       504 & 183.9797 & -13.0210 &            100.0 &        \cite{Heintz2020} \\
FRB20191001A &   0.2340 &       507 & 323.3517 & -54.7483 &            100.0 &        \cite{Heintz2020}\\
FRB20200430A &   0.1610 &       380 & 229.7065 &  12.3763 &            100.0 &        \cite{Heintz2020} \\
FRB20200906A &   0.3688 &       578 &  53.4955 & -14.0830 &            100.0 &      \cite{bhandari2021} \\
FRB20210117A &   0.2145 &       730 & 339.9792 & -16.1514 &             99.8 &     \cite{bhandari2023a} \\
FRB20210320C &   0.2797 &       385 & 204.4587 & -16.1227 &            100.0 &          \cite{james+23} \\
FRB20210410D &   0.1415 &       575 & 326.0862 & -79.3182 &             99.6 &             \cite{caleb} \\
\hline
\end{tabular}

\caption{Sample of FRBs. (1) Name of the FRB; (2) Host redshift; (3) Total dispersion measure observed; (4) and (5) Right Ascension and Declination of the FRB (J2000), respectively; (6) PATH posterior probability of the putative host; (7) Discovery paper reference for each FRB; \rf{note that} some values \rf{reported in this table} have been updated \rf{according to} \cite{gordon23} and \cite{shannonICS}.} \label{tab:sample}
\end{table*}

\subsection{Previous observations}
\label{sec:obs}
The Fast and Fortunate for FRB Follow-up (F$^4$) is a collaboration endeavoring to obtain dedicated photometric and spectroscopic follow-up observations of localized FRBs and their host galaxies. All the observational data of our studied host galaxies except \halpha\ fluxes (e.g. stellar masses, SFRs, FRB offsets), come from previous work done by F$^4$ (see last column of Table~\ref{tab:sample} for specific references) and are available on both the \href{https://github.com/FRBs/FRB}{FRB GitHub repository}\footnote{https://github.com/FRBs/FRB} (\citealp{FFFF})
and the `FRB Hosts' website\footnote{https://www.frb-hosts.org/}. We note that some values have been updated since their discovery papers, and here we used those reported by \citet{gordon23} \rf{and \citet{shannonICS}.}

\subsection{VLT/MUSE observations and data reduction}

We wish to study the \dmhost\ empirical contribution, and for this we use the Multi Unit Spectroscopic Explorer (MUSE; \citealt{MUSE}) mounted on UT4 (Yepun) at the Very Large Telescope (VLT) on Cerro Paranal, Chile. MUSE provides resolved or partially resolved integral-field spectroscopy which allows for a `local' estimation of the ISM contribution to the host \dmhism\ based on the \halpha\ nebular emission at the FRB position (when available; see Section~\ref{sec:empiric}),
and a `global' estimation based on the \halpha\ nebular emission of the entire galaxy.

All the MUSE observations were conducted with the wide field mode (WFM) with adaptive optics and nominal wavelength range (AO-N) mode. This setup provides a field of view (FoV) of $1\arcmin \times 1\arcmin$ with a spatial sampling of 0.2\arcsec\,pix$^{-1}$, and spectral coverage of $\approx 4700-9300$\footnote{With a gap between $5760-6010$\,\AA\ due to the contamination produced by the Sodium laser used for AO.}\AA, with resolving power ranging from \textit{R}$\simeq$1770 at 4800\,\AA\ to \textit{R}$\simeq$3590 at 9300\,\AA. We observe each field for $\sim 4-8 \times 600$\,s individual exposures with typical DIMM seeing conditions of $\approx 1\arcsec$, which translated to $\lesssim 0.8$\arcsec\ PSF thanks to the enabled AO. Table~\ref{tab:muse} summarizes the dates, exposure times and program IDs of the observations.

Data reduction was performed using the ESO MUSE pipeline \citep[v2.8.6,]{Weilbacher2020} within the ESO Recipe Execution Tool (EsoRex) environment \citep{ESOREX2015}. We used standard procedure and parameters.

\begin{table*}[ht]
\centering
\begin{tabular}{cccc}
\hline
\hline
    FRB name & Obs. Dates (UT) &   Exp. time (s) &   Program ID \\
(1)&(2)&(3) & (4)\\ 
\hline
FRB20180924B & 2018 Nov 05 / 2019 Dec 06  &  $8\times 600$  &  2102.A-5005(A) / 104.A-0411(A)\\
% & 2019 Dec 06& & XX\\
FRB20190102C &     2022 Oct 15  &  $4\times 600$  &  110.241Y.001 \\
FRB20190608B &    2022 July 29  &  $8\times 600$  &  105.20HG.001 \\
FRB20190611B &    2021 July 02  &  $7\times 600$  &  105.20HG.001 \\
FRB20190711A & 2021 July 02 / 2021 Ago 10 &  $8\times 600$  &  105.20HG.001 \\
FRB20190714A & 2021 Apr 18 / 2022 Feb 05  &  $8\times 600$  &  105.20HG.001 \\
FRB20191001A &   2022 Oct 1 / 2022 Oct 16  &  $8\times 600$  &  110.241Y.001 \\
FRB20200430A &     2023 Mar 31  &  $8\times 600$  &  110.241Y.002 \\
FRB20200906A &  2022 Nov 26/ 2022 Nov 29  &  $8\times 600$  &  110.241Y.002 \\
FRB20210117A &     2022 Oct 03  &  $4\times 600$  &  110.241Y.001 \\
FRB20210320C & 2023 May 11 / 2023 Jun 23  &  $8\times 600$  &  110.241Y.002 \\
FRB20210410D &  2022 Oct 22 / 2022 Oct 25  &  $8\times 600$  &  110.241Y.001 \\
\hline
\end{tabular}
\caption{Summary of MUSE observations. (1) Name of the FRB; (2) Date of the observations; (3) Exposure time; (4) Program ID(s).}
\label{tab:muse}
\end{table*}

%-----------------------------------------------------------------
\section{Analysis}
\label{sec:analysis}

\subsection{Direct \dmhost\ estimation, \dmhdirect}
\label{sec:empiric}

Here we provide a direct estimate of \dmhdirect. 
For this estimation we consider two contributions,
 \dmhism\ and \dmhhalo, giving 
\begin{equation}
    \dmhdirect = \dmhism + \dmhhalo ,
    \label{eq:dmhost}
\end{equation}
where \dmhism\ includes gas from the ISM of the galaxy host plus any possible gas associated to the FRB progenitor itself,
while \dmhhalo\ is the contribution from the galactic halo, both defined in the source frame.
The methodology to estimate these two contributions is presented as follows.

\subsubsection{Estimating \dmhism}
\label{sec:dmism}

To estimate \dmhism\ for the different galaxy hosts, we adopt the procedure outlined by \citet{Tendulkar2017} 
following \cite{reynolds1977}. 
For completeness, we include the key equations here.

From the \halpha\ flux measured by MUSE, we obtain the H$\alpha$ surface brightness, \sbha , by dividing the \halpha\ line emission by the effective area considered. 

We correct the emission for the Galactic dust extinction (\citealp{fm07}) and by surface brightness dimming in order to pass from the observer's frame to the source's frame. From \sbha, in the source's frame, we use equation 10 of \cite{reynolds1977} to obtain an estimate of the emission measure (EM):

\begin{equation}
    {\rm EM}(\halpha) = 2.75 \text{ pc cm} ^{-6} \text{ T}_{4}^{0.9} \left[\frac{\sbha}{\rm{Rayleigh}}\right] \ \rm{.}
    \label{eq:em}
\end{equation}

\noindent Then, we adopt equation~5 from \cite{Tendulkar2017} to estimate \dmhism, in the source's frame,

\begin{equation}
    \dmhism =387 \text{ pc cm}^{-3} L^{1/2}_{kpc} \left[ \frac{f_f}{\tau(1+\zeta^2)/4}      \right]^{1/2} 
    \left( \frac{\rm{EM}}{600 \text{ pc cm} ^{-6} }   \right)^{1/2} \ \rm{,}
    \label{eq:dmhism}
\end{equation}
where $f_f$ is the volume filling-factor of clouds, $\zeta$ represents the fractional density variation within any given cloud of ionized gas due to turbulence inside clouds, and $\tau$ is the density variation between any two clouds, all over a path length $L_{\rm kpc}$ in kpc (\citealp{Cordes_chittidi, Tendulkar2017,ChittidiFRB190608}).

\rf{Equation~\ref{eq:dmhism}  has been calibrated in the Milky Way and may not necessarily apply to FRB hosts in general. More fundamentally, even within the Milky Way, the clumpiness and density variations can vary significantly between sightlines and thus this estimation is inherently uncertain. Therefore, the exact values of the parameters governing \dmhism\ are unknown. } In what follows we assume a turbulent and dense environment in the sightline, that is, the volume filling-factor is maximum ($f_f$=1) and that each ionized cloud along the sightline has internal density variations dominated by turbulence ($\zeta$=1) and that the variation between clouds is maximum ($\tau$=2). Finally, we assume that the FRB comes from a path width similar to the Milky Way disk, that is, L$_{\rm kpc}$=0.15.  In Section~\ref{sec:systematic} we quantify the effect of these systematic uncertainties \rf{on} our overall results.

\subsubsection{Global and local \dmhism}
\label{sec:gvsl}
%global estimate

To provide a global estimation for \dmhism, we use the \halpha\ nebular emission within the area defined by an ellipse with the reported eccentricity and by using the galaxy's effective radius as the semi-major axis, referred to as \dmhiglobal.

\begin{figure}[ht]
\centering
\includegraphics[width=0.4\textwidth]{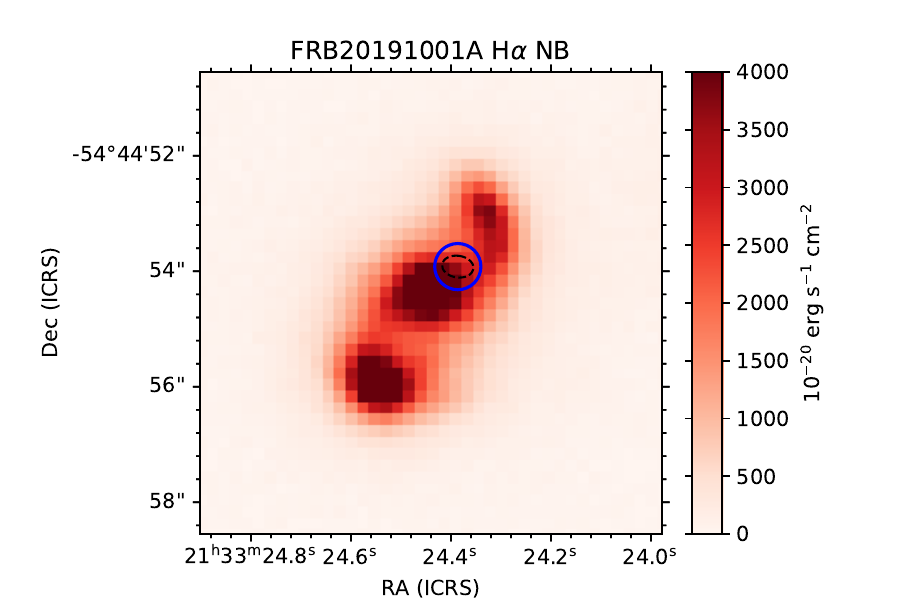}
\includegraphics[width=0.4\textwidth]{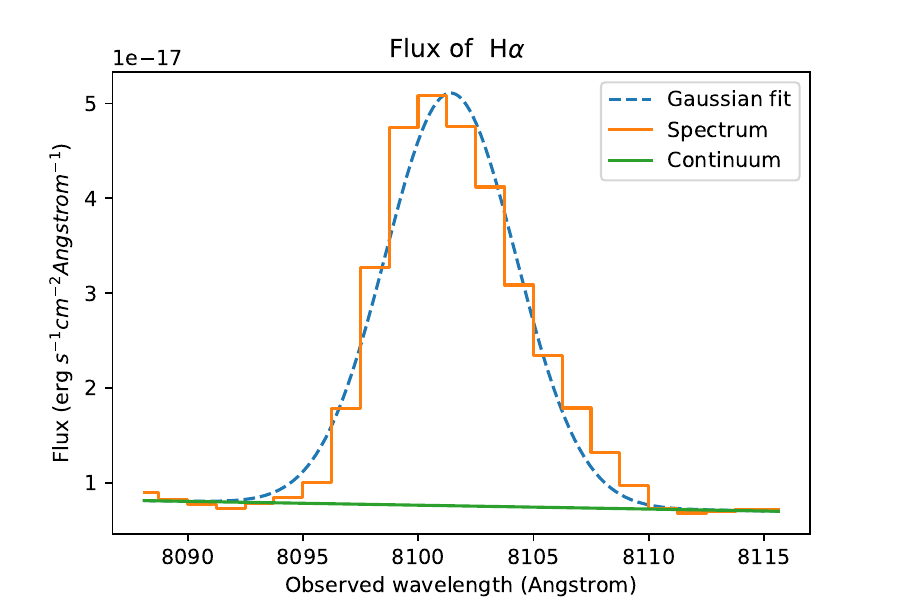}
\caption{Upper: FRB20191001A host emission integrated at the narrow band (NB) encompassing \halpha\ using the same wavelength range as shown in the bottom panel. The blue circle is centered at the FRB localization and has a radius of $0.4$\arcsec\ and the dashed black ellipse represents the actual FRB position uncertainty. Lower: The integrated spectrum of the \halpha\ emission within the blue circle (orange). We fit this emission with a Gaussian (dashed blue) plus a continuum (green).}
\label{fig:halpha}
\end{figure}

%local estimate
In addition, to have a more relevant \dmhism\ estimation, we use the local nebular emission of H$\alpha$ coincident with the FRB position, referred to as \dmhilocal. We obtain the local H$\alpha$ emission from the MUSE datacubes, by selecting a region of radius $0.4$\,arcsec ($2$ native spaxels) which corresponds to the typical seeing (comparable to or larger than the typical FRB positional uncertainty in our sample). Then, from this region we extract the spectrum at the observed wavelengths corresponding to the \halpha\ transition by summing the flux of all the spaxels with equal weight. We model the continuum by a spline which is subtracted from the observed spectrum. Then we model the remaining emission line flux with a Gaussian fit.
Figure~\ref{fig:halpha} shows an example for FRB20191001A. The FRB localization is denoted by the blue circle. In the lower panel of the figure we have the extracted 1D spectrum within the region, at the wavelengths corresponding to the observed H$\alpha$ emission, together with our modeled Gaussian fit and continuum. Finally we integrate the Gaussian fit to obtain the total \halpha\ nebular emission flux and its error. 

Table~\ref{tab:localvsglobal} summarizes our results obtained for the global and local \dmhism\ estimates, and Figures~\ref{fig:halpha1}~to~\ref{fig:halpha12} in the Appendix show the \halpha\ maps and emission lines for the rest of the sample.

Uncertainties in these estimations are calculated as follows: 
We do $10^5$ realization of \halpha, following a normal distribution around the given/calculated \halpha\ for every FRB host, with the error of \halpha\ as standard deviation; then, with every realization we repeat the previous steps to provide an empirical 1$\sigma$ uncertainty from the resulting PDF.

For the host galaxy of FRB20190711A at $z=0.522$, the MUSE spectrum does not cover the redshifted \halpha\ wavelength. In this case we follow a procedure similar to that used by \citealp{Logrono}, \citealp{Calzetti2000} and \citealp{ChittidiFRB190608}.
We first obtained the \hbeta\ flux and \hgamma\ analogously. We then estimate the extinction from the observed ratio of H$\gamma$ to H$\beta$ flux ($0.397$), and by comparing it to the theoretical value (0.466, \citealt{Osterbrock}). This gives us an extinction estimate of A$_{v}=$ 0.18. Then with the extinction we can calculate the intrinsic ratio of H$\alpha$ to H$\beta$ flux following \citealt{Calzetti2000, Logrono} to obtain a ratio of $4.364$. Finally, this ratio gives us our estimate of H$\alpha$ (corrected for dust).

\begin{figure}[ht]
 \centering
    \includegraphics[width=0.4\textwidth]{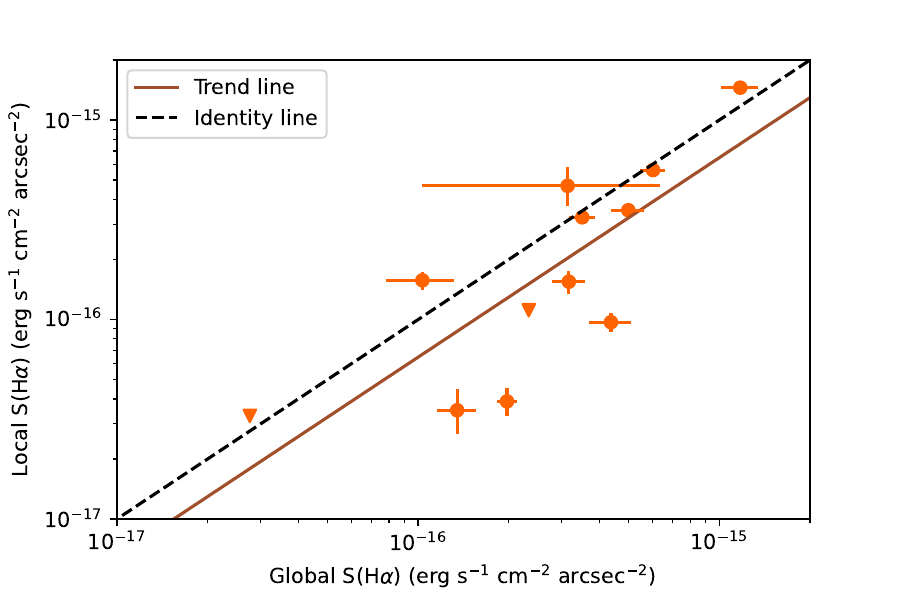}
    \caption{Comparison between the local and global inferred surface brightness of \halpha, obtained from MUSE cubes, for our sample. The trend line (dark orange) has a slope of 1 \rf{by construction. The identity line (dashed line) corresponds to the 1:1 relation}.}
 \label{fig:local_vs_global}
\end{figure}

Given Equations~\ref{eq:em}~and~\ref{eq:dmhism}, besides geometrical factors and temperature, the key observable for inferring \dmhism\ is the \sbha. In our case we obtain \sbha\ for global (full galaxy) and local (at the FRB position) \halpha, and a comparison between the two is shown in Figure~\ref{fig:local_vs_global}. 
We observe that the global measurement is systematically larger than the local one. The dark-orange line shows the linear correlation between them, from which we infer a systematic difference of \dha\%, which translates to $\approx$\ddm\% in the inferred \dmhism\ (see Equations~\ref{eq:em}~and~\ref{eq:dmhism}).

\subsubsection{\dmhhalo}
\label{sec:dmhhalo}

For estimating \dmhhalo\ we follow the procedure similar to \citealp{ProchaskaZheng2019}, \citealp{simha+23} and \citealp{ChittidiFRB190608}\rf{.} 
From the published estimated stellar mass for each host galaxy, we implement the abundance matching technique to infer a halo mass (\citealp{Moster2013}).
Then, following \cite{ProchaskaZheng2019}, we can estimate \dmhhalo\ assuming a density profile for the halo gas.
We used the modified Navarro, Frenk \& White (mNFW) profile described by \cite{ProchaskaZheng2019} 
following \cite{mathews1X},

\begin{equation}
\rho_b=\frac{\rho_b^0}{y^{1-\alpha}(y_0+y)^{2+\alpha}}
\end{equation}

\noindent where $y_0$ = 2 and $\alpha$ = 2, $y \equiv c (r/r_{200})$, with $c$ being the concentration parameter and $r_{200}$ is defined as the radius within which the average density is 200 times the critical density. Because we expect that not all of this gas is ionized, we need to estimate the fraction of baryons that are ionized ($f_{hot}$). For this, we use $f_{hot}=55\%$ \citep{khrykin2024a} as a fiducial value to calculate $\rho_b^0$, which corresponds to the density of the ionized mass.

With $\rho_b$, we can calculate the density of free electrons, $n_e$ as:

\begin{equation}
    n_e=\mu _e \frac{\rho_b}{m_p \mu_H}
\end{equation}

\noindent with $m_p$ the proton mass, $\mu_H$ = 1.3 the reduced mass (accounting for Helium) and $\mu_e$ = 1.167 accounts for fully ionized Helium and Hydrogen. Corrections for heavy elements are negligible.

This profile is integrated along the line-of-sight, from the projected distance $R$ of the FRB, to the max radius of the halo (r$_{200}$).
The DM$_{\rm halo}(R)$ value is defined as:

\begin{equation}
    \dmhhalo(R)=\int\limits^{\sqrt{r_{200}^2-R^2}}_0 n_e \rm{d} s
    \label{eq:pz19}
\end{equation}

\noindent where the lower integration limit corresponds to the mid-plane of the halo, which is perpendicular to the line of sight. Finally we evaluate Equation~\ref{eq:pz19} to obtain 
\dmhhalo, in the source's frame.

\begin{figure}[ht]
\centering
\includegraphics[width=0.4\textwidth]{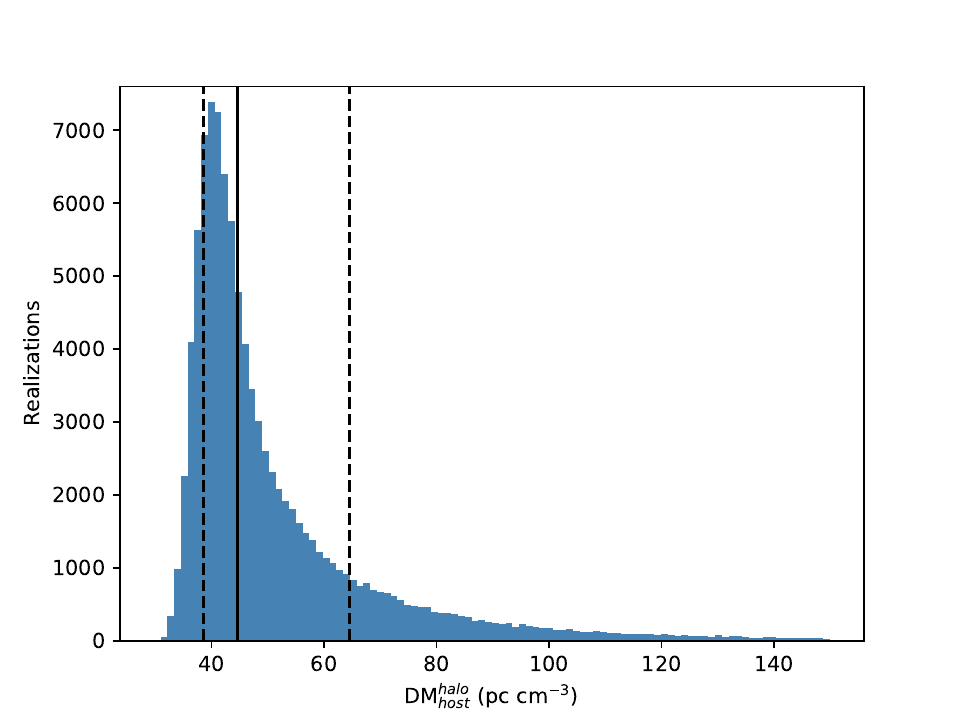}
\caption{Distribution of \dmhhalo\ for FRB20191001A. The vertical black solid line represents the median value and the dashed black lines the corresponding 16th and 84th percentiles.}
\label{fig:PDF}
\end{figure}

To account for uncertainties in \dmhhalo\ estimations derived from stellar mass for each host galaxy, we generate $1\,000$ realizations of stellar masses, assuming a log-normal distribution where the standard deviation corresponds to the stellar mass uncertainty. 
For each stellar mass realization, we generate $100$ additional realizations, varying the $8$ parameters of the Moster relation. The mean values and standard deviations for these parameters are taken from table~1 of \cite{Moster2013}, following a normal distribution. Finally, we obtain a PDF of \dmhhalo\ with $10^{5}$ realizations per FRB, and use the $1\sigma$ of this PDF as our final uncertainty. As an example, Figure \ref{fig:PDF} shows this PDF for FRB20191001A.

\subsubsection{Estimating \dmhdirect}%DM$_{\rm host, empirical}$}
\label{sec:dmdirect}

To obtain \dmhdirect\  \rf{we add the two PDFs of \dmhhalo\ and \dmhism}. From this sum, we generate a new PDF of \dmhdirect\ for each galaxy host in our sample, from which we use the median value as our \dmhdirect\ estimation, and $1\sigma$ for its uncertainties.
For the calculation of \dmhdirect\ we use 
\dmhilocal. \rf{We note that in two cases (FRB20190611B and FRB20210117), the local \halpha\ emission line was not detected (upper limits in Table~\ref{tab:localvsglobal}). However, these upper limits imply \dmhism\ values comparable with those inferred from the global measurements (Figure~\ref{fig:local_vs_global}). Given that our systematic uncertainties in \dmhism\ are much larger than these differences (see below), for simplicity in the following we treat these upper limits as actual measurements.}

The systematic uncertainties of changing the values of the parameters in Equation~\ref{eq:dmhism} are not taken into account for our reported values. For reference, these can make the \dmhism\ up to 2 (3) times larger (smaller) than our fiducial values (see also Section~\ref{sec:systematic}). Similarly, the systematic uncertainty from the chosen path length, either in the ISM or in the halo, can make \dmhism/\dmhhalo\ up to 1.4 or 2 times larger respectively, while in the case of the FRB being in the outskirts of the halo (unlikely given the known projected distance distributions), both these contributions could be as small as zero.

\begin{figure*}[ht]
    \centering
    \includegraphics[width=0.9\textwidth]{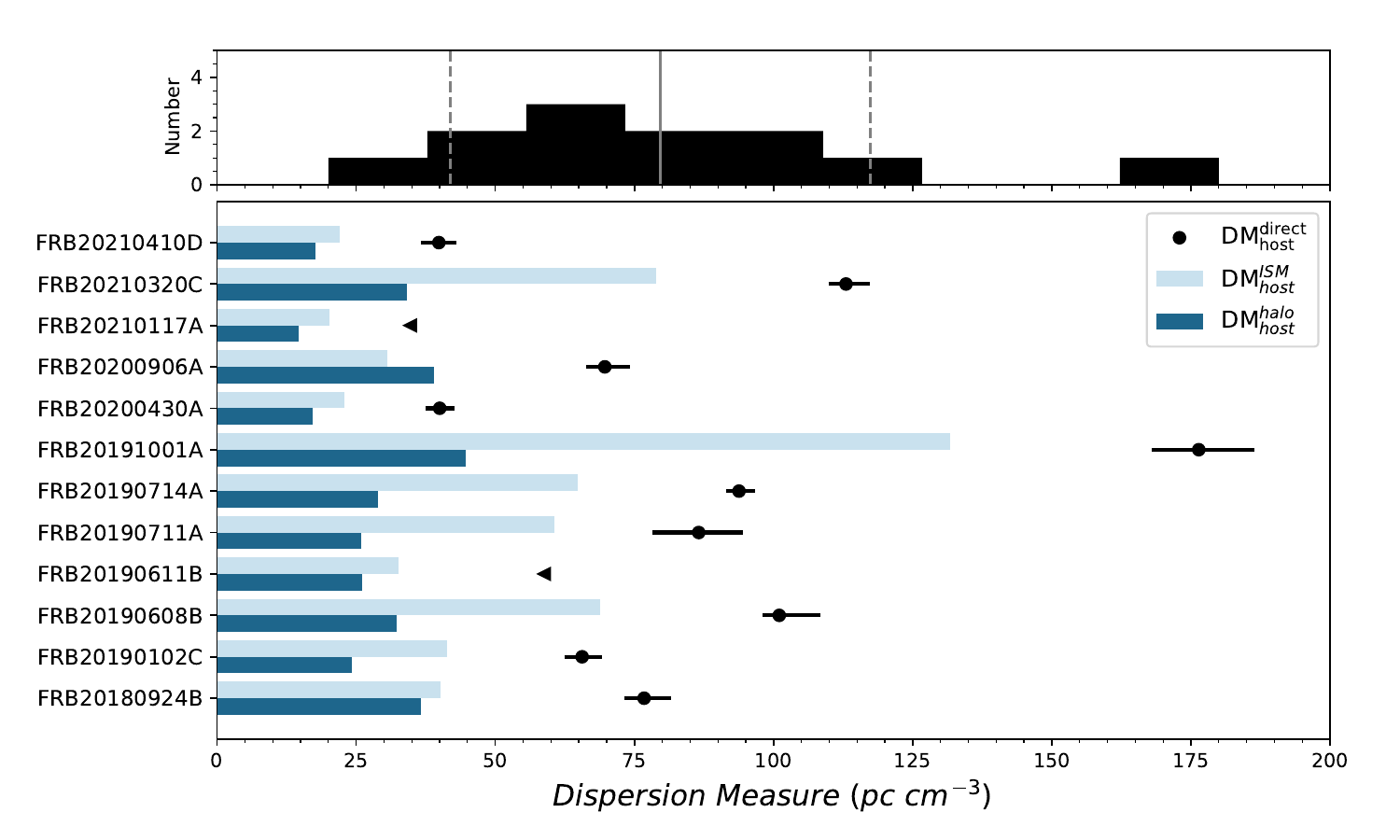}
    \caption{
    Estimates of \dmhost\ for the \nsamp\ FRB hosts, which shows our empirical direct estimate \dmhdirect\ (black points). The main panel shows each contribution of \dmhdirect\ , \dmhhalo\ in dark blue bars and \dmhism\ in light-blue bars. 
    %Estimates of \dmhost\ for our \nsamp\ galaxy hosts, where the main panel shows each component of DMdirect of each FRB, and in black points dm direct with its unciertities. 
    The upper panel is the histogram for \dmhdirect\ \rf{where the mean and the standard deviation ($\pm 1\sigma$) of the distribution are represented by the solid and dashed vertical lines, respectively}.}
    \label{fig:sample}
\end{figure*}

\subsection{Estimating \dmhost\ 
with the Macquart Relation}
\label{sec:macquart}

For comparison, we also provide an indirect estimation of \dmhost\, referred to as \dmhmacq\ (rest-frame) using the follow equation:

\begin{equation}
\dmhmacq = (1+z)(\dmfrb - \dmmw - \dmacosmic)
\label{eq:host}
\end{equation} 

\noindent where \dmacosmic\ is the average contribution of extragalactic gas at redshift $z$. This analysis requires estimations of \dmmw\ and 
\dmcosmic.

% %%%%%%%%%%%%%%%%%%%%%%%%%%%%%%%%%%%%%%%%%%%%%%%%%%%
\subsubsection{DM$_{\rm MW}$ contribution}

The Milky Way contribution comes from the ISM and the halo (Equation~\ref{eq:mw}). For \dmmwism,  we use the NE2001 model \citep{NE2001-II}, where the free electrons are integrated across the Galaxy according to the FRB coordinates. NE2001 is a Galactic distribution model of free electrons which actually comprises two disk components: spiral arms, and localized regions \rf{(like the Magellanic Clouds)}. For \dmmwhalo, we use a fixed value of 40~pc\,cm$^{-3}$ \citep{ProchaskaZheng2019}.

\subsubsection{DM$_{\rm host}^{\rm Macquart}$ and upper limit on DM$_{\rm host}$}

Given our estimates of \dmmw\ for each FRB sightline, and \dmacosmic\ for an average sightline up to redshift $z=z_{\rm frb}$ (given by $\Lambda$CDM), we use Equation~\ref{eq:host} to obtain the predicted Macquart \dmhost, \dmhmacq. We note that for two cases this \dmhmacq\ results in a negative (non-physical) value; we nevertheless report them as such for the sake of completeness and for statistical consistency (see Section~\ref{sec:discussion}).

We caution the reader that every sightline is unique, and thus we expect deviations from \dmacosmic\ for individual FRBs with the majority of \dmcosmic\
expected to lie below this value 
\citep[e.g.][]{baptista+2024}. This in turn implies that our \dmhmacq\ value will most of the time be an underestimation of the true \dmhost, while the rest of the time an overestimation. However, we expect that the average \dmhmacq\ value of the ensemble of measurements to be close to the average true \dmhost\ value of the ensemble, modulo systematic errors e.g.\ in \dmmw.

We use the same principle to estimate a maximum possible value of \dmhost\ (rest-frame), referred to as \dmhmax, for a given individual FRB, following:

\begin{equation}
\dmhmax= (1+z)(\dmfrb - \dmmwmin - \dmcosmicmin) \ ,
\label{eq:host_max}
\end{equation}

\noindent where \dmmwmin\ is the minimum possible value for \dmmw\ allowed by the NE2001 model, and \dmcosmicmin\ being a conservative minimum contribution to \dmfrb\ by the IGM. For this, we use the lower $1\%$ percentile of the allowed scatter around the Macquart relation, that is, assuming that the FRB traveled through an extremely under-dense sightline, without intersecting any dense cosmic web filament nor intervening halos of galaxies. Although it is very unlikely to observe such an extreme situation in a sample of only \nsamp\ FRBs, we use this estimation to provide a conservative upper limit for the \dmhost.

%-----------------------------------------------------------------
%\section{Results}
%\label{sec:results}

\begin{figure*}[ht]
\centering
\includegraphics[width=0.9\textwidth]{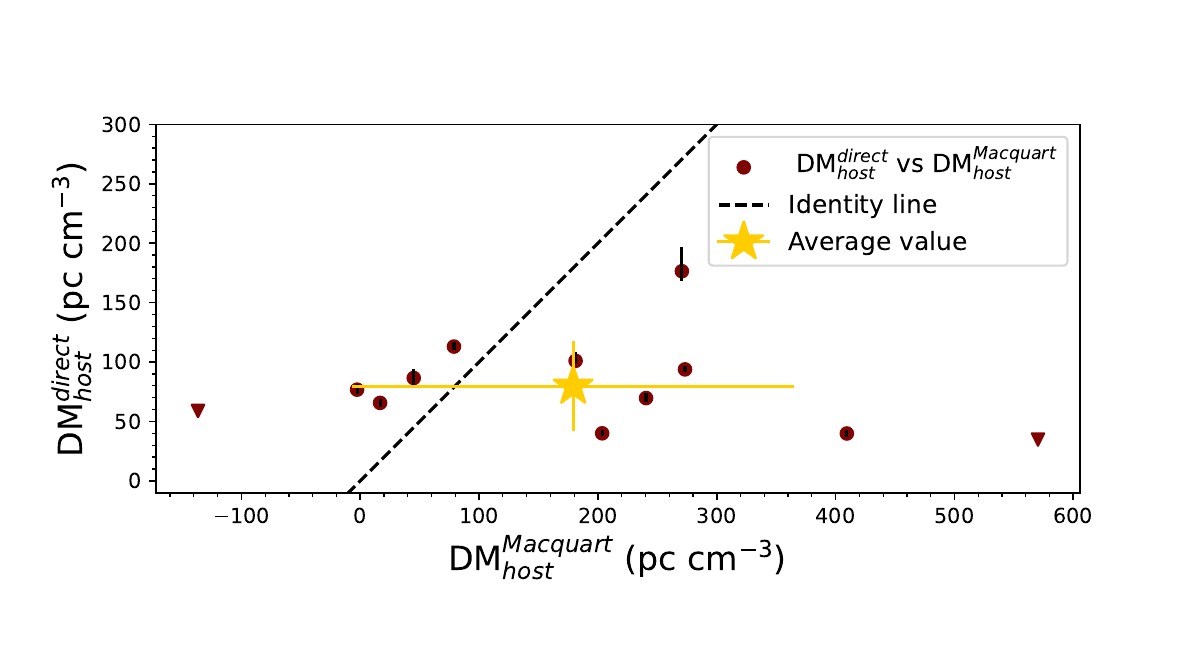}
\caption{Relation between \dmhdirect\ and \dmhmacq\ (dark points), with error bars showing the statistical uncertainties of \dmhdirect\ (we do not estimate uncertainties of our \dmhmacq\ values). The dashed line corresponds to the identity line (i.e. the 1:1 relation). The golden star corresponds to the average values of \dmhdirect\ and \dmhmacq\ with the error \rf{bars} being their corresponding standard deviation.}
\label{fig:empiric_vs_macquart}
\end{figure*}

\begin{figure*}
    \includegraphics[width=0.33\textwidth]{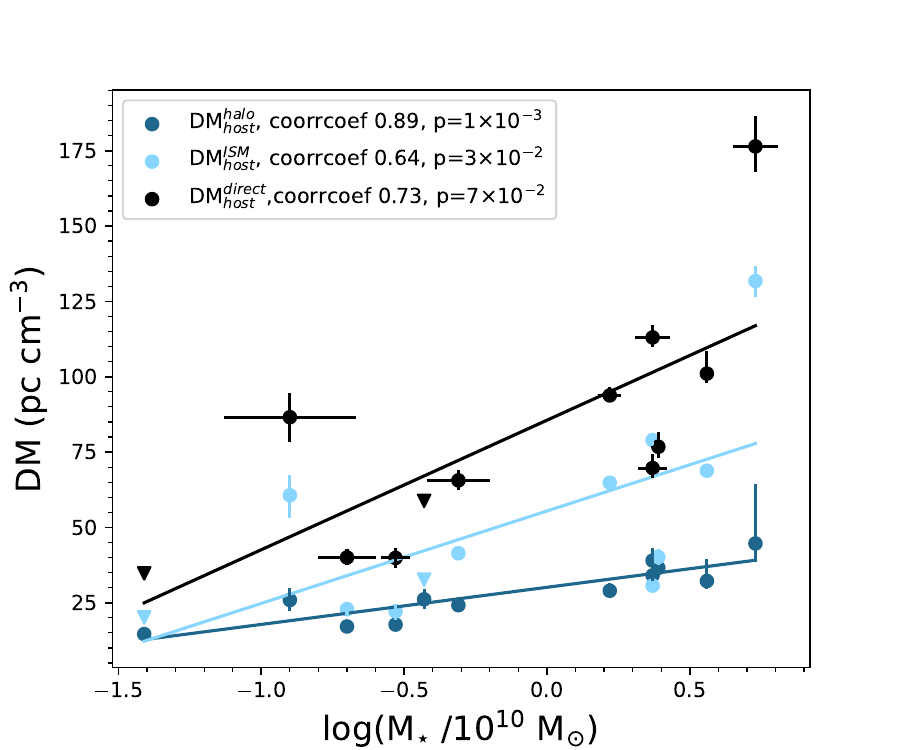}
    \includegraphics[width=0.33\textwidth]{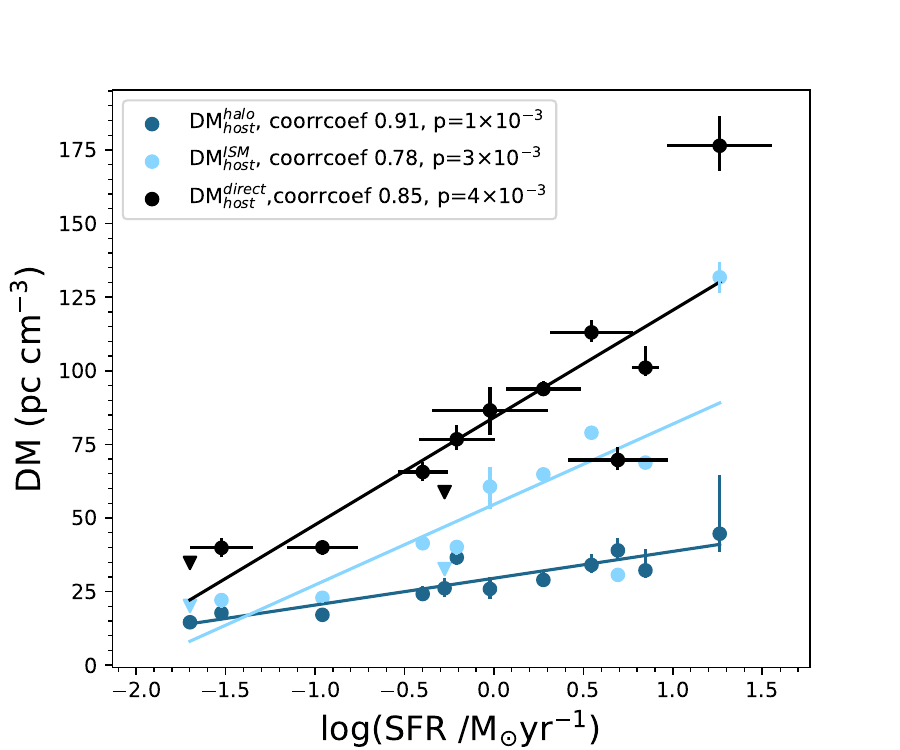}
    \includegraphics[width=0.33\textwidth]{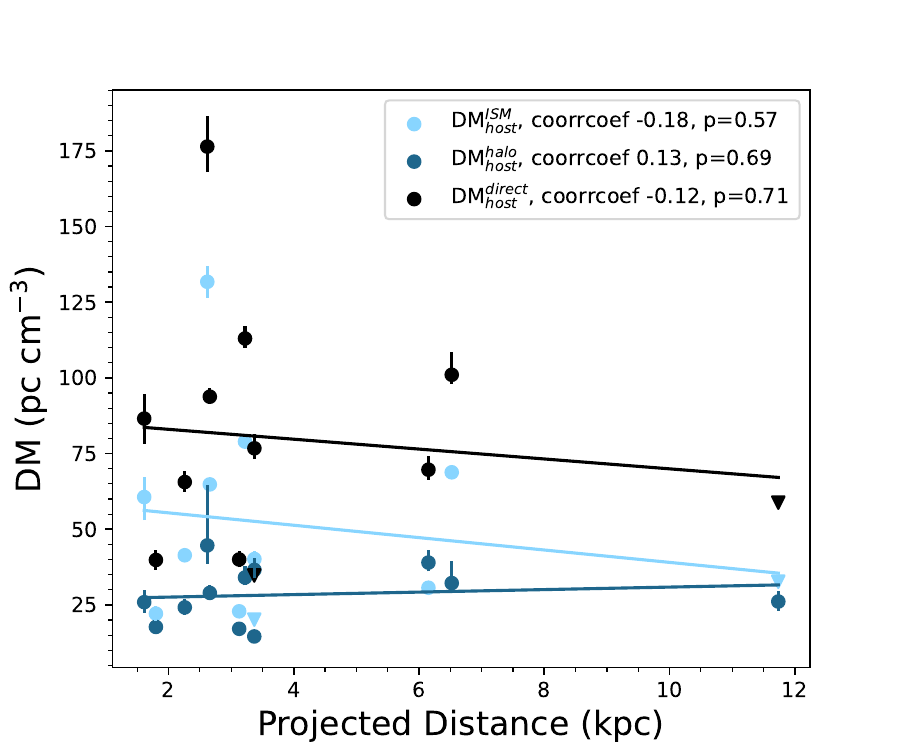}
    \caption{Rest-frame dispersion measure (DM) as a function of stellar mass of the host (left), star-formation rate of the host (SFR; middle), and projected offset from the center of the host (right).
    %Left: DM$_{\rm host}$ vs stellar mass. Center: DM$_{\rm host}$ vs SFR. Right: DM$_{\rm host}$ as a function of offset by center of galaxy. 
    The black points correspond to \dmhdirect, while the light-blue and the dark-blue points correspond to \dmhism\ and \dmhhalo, respectively. The black line shows a linear fit for \dmhdirect\ (= \dmhhalo\ + \dmhism), while
    the light-blue and dark-blue lines are the linear fits of \dmhism\ and \dmhhalo, respectively. The parameters of these fits are presented in Table~\ref{tab:params}, and \rf{the Pearson coefficients and $p$-values are also given in the figure legends.}
    } \label{fig:mass_sfr_offset}
\end{figure*}

\begin{figure}[ht]

\centering
\includegraphics[width=0.49\textwidth]{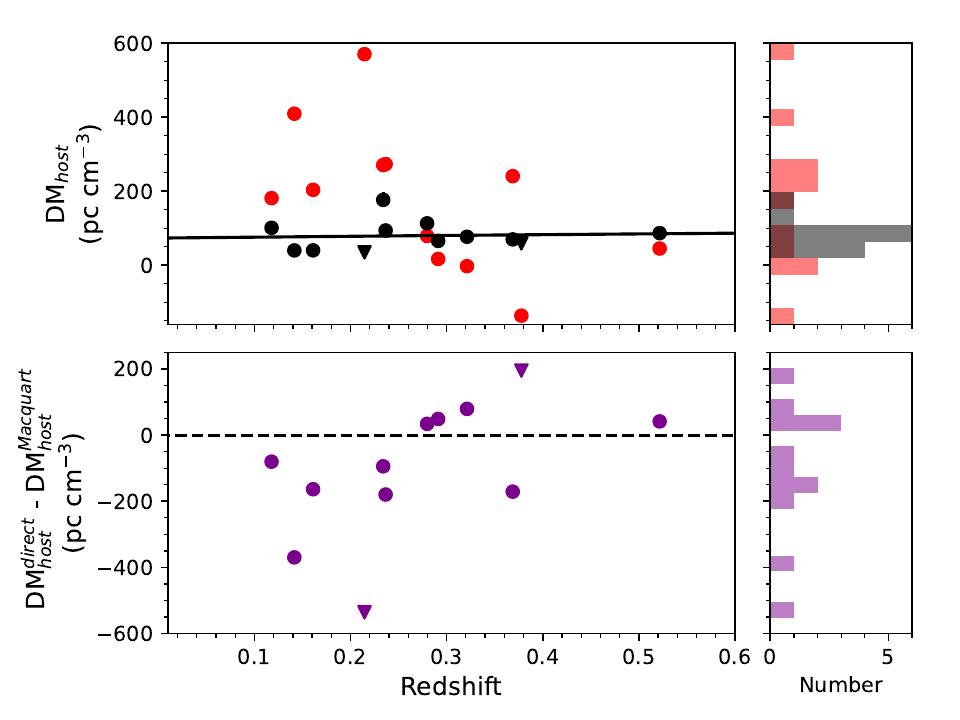}
\caption{Top panel: \dmhost\ as a function of redshift, black and red points correspond to \dmhdirect\ and \dmhmacq, respectively. The black line is the tendency of \dmhdirect. Negative values for \dmhmacq\ are nonphysical; we nevertheless report them as such for statistical consistency. Histograms of the different samples are shown in the right respective panels following the same color scheme. Bottom panel: The purple points show the difference between \dmhdirect\ and \dmhmacq\ as a function of redshift. The black dashed line at value 0 indicates where \dmhdirect\ and \dmhmacq\ are equal. Histogram of the sample is shown in the right panel.}
\label{fig:redshift}
\end{figure}

 %\caption{Left: DM$_{\rm host}$ as a function of redshift, blue and green points are calculates empirically and by Macquart relation, respectively. The blue line is the tendency of the blue points. In difference with figure 2 here, negative values for DM$_{\rm host,macquart}$, conserve its calculate value, instead set it in 1.  Right: This points show the difference between DM$_{\rm host,macquart}$ and DM$_{\rm host,empiric}$ in function of redshift, the black dashed line mark represent value 0.}

%    \includegraphics[width=0.4\textwidth]{data/dm_vs_z_histogram.pdf}}

%-----------------------------------------------------------------
\section{Results and Discussion}
\label{sec:discussion}

Our results on \dmhdirect\ are shown in Figure~\ref{fig:sample} and \rf{summarized} in Tables~\ref{tab:localvsglobal}~and~\ref{tab:summary}. \rf{We have found that on average $\langle$DM$_{\rm host} \rangle = \avgdirect$\,\pccc\ with a standard deviation of the \dmhdirect\ distribution of $\stddirect$\,\pccc. This value} is larger than the typically used value of $50$\,\pccc\ (\citealp{Arcus2020}) but consistent within uncertainties. In the following we investigate possible correlations of the individual \dmhdirect\ values as a function of galaxy properties including redshift, and also investigate the impact of possible systematic biases present in our current estimates.

\subsection{Comparison between \dmhdirect\ and \dmhmacq}
\label{sec:comparison}

Figure~\ref{fig:empiric_vs_macquart} shows a comparison between our \dmhdirect\ measurements and the \dmhost\ expectation from the Macquart relation, \dmhmacq\ (dark points). The figure also shows their corresponding average values of the ensemble (golden star):
$ \langle \dmhmacq \rangle = 179 \pm 54$\
\pccc, and 
$\langle \dmhdirect \rangle = \avgdirect$\ 
\pccc, \rf{with the error bars being their corresponding standard deviations ($186$ and $31$\,\pccc, respectively).}

Although individual measurements differ significantly between \dmhdirect\ and \dmhmacq\ (as expected), the average value is consistent between them, given their statistical uncertainties. We observe that 5(7) out of 12 of our \dmhdirect\ measurements are larger(smaller) than their corresponding \dmhmacq\ values, implying that no obvious bias is present in our sample of FRBs and corresponding host galaxies for estimating the average \dmhdirect\ (see Section~\ref{sec:macquart}). Furthermore, we observe that the scatter in $\langle \dmhdirect \rangle$ is about $\sim 5 \times$ smaller than the scatter in $\langle \dmhmacq \rangle$ which indicates that \dmhdirect\ may be a more precise measurement compared to \dmhmacq; however, our empirical $\langle \dmhdirect \rangle$ estimate may be subject to systematic uncertainties (of order $30\%$; see Section~\ref{sec:systematic}) making it (in principle) less accurate than $ \langle \dmhmacq \rangle$. Indeed, given that mostly cosmological parameters are involved in the estimation of \dmhmacq, \rf{we only expect possible biases due to systematic errors in our individual \dmmw\ estimates or from the stochastic effect of foreground massive halos that could contribute to the observed DM \citep{lee+23}.}
Although $\langle \dmhmacq \rangle$ may be more accurate it is nevertheless highly uncertain, but the fact that we find consistency between  $\langle \dmhdirect \rangle$ and $ \langle \dmhmacq \rangle$ may indicate that systematic uncertainties are not dominating the measurements. Moreover, we have computed the maximum possible \dmhmax\ given our cosmological parameters (see Section~\ref{sec:macquart}) and all our \dmhdirect\ measurements are smaller than such a limit (see the fifth and sixth columns in Table~\ref{tab:summary}), ensuring consistency.

Finally, we note that we do not see any strong correlation between \dmhdirect\ and \dmhmacq\ (Pearson coefficient of $-0.2$, \rf{with $p$-value of 0.6}) indicating that our current modeling could be missing an important DM contribution to the \dmfrb, i.e., some FRBs may have large intrinsic \dmhost\ and/or a much larger actual \dmcosmic\ compared to the \dmacosmic\ (e.g. due to intervening foreground structures) that our modeling is not capturing or taken in consideration. Indeed, the apparent lack of correlation between the two estimates is partially driven by the two data-points with the largest \dmhmacq.

\subsection{Correlations between \dmhdirect\ and galaxy properties}
\label{sec:correlations}
The left panel of Figure~\ref{fig:mass_sfr_offset} compares the \dmhdirect\ (black points) to the corresponding stellar masses of the host galaxies. We also show the individual contributions to the \dmhdirect\ from \dmhhalo\ (dark blue) and \dmhism\ (light blue). We observe a positive correlation between \dmhdirect\ and stellar mass, of the form:

\begin{equation}
    \left(\frac{\dmhost}{\rm{ pc\,cm}^{-3}}\right) \approx 43 \,\log{\left(\frac{M_{\star}}{10^{10}M_{\odot}}\right)} + 86 
    \label{eq:dmh_mass}
\end{equation}

\noindent with a Pearson correlation test giving a coefficient of $0.73$ (with $p$-value of 7 $\times$ $10^{-2}$). 
% next sentences could go into discussion instead.
A correlation of \dmhost\ with stellar mass is expected, given that the larger the stellar mass, the larger the halo mass used to estimate \dmhhalo\ (see Section~\ref{sec:dmhhalo}). Indeed, we also see a positive correlation between \dmhhalo\  and stellar mass with Pearson coefficient of $0.89$ (with $p$-value of 1 $\times$ $10^{-3}$). Interestingly, a positive correlation is also present between \dmhism\ and stellar mass (Pearson coefficient of $0.64$, with $p$-value of 3 $\times$ $10^{-2}$), which we attribute to the fact that most galaxies in our sample are star-forming and the existence of the star-formation main sequence (SFMS; e.g. \citealt{sfr_first}; see below).

The middle panel of Figure~\ref{fig:mass_sfr_offset} shows \dmhdirect\ (black points) as a function of global SFR for each host galaxy; as before, the individual contributions of \dmhhalo\ (dark blue) and \dmhism\ (light blue) are also shown. In this case we also observe a positive correlation between \dmhdirect\ (and its components) and global SFR of the form:

\begin{equation}
    \left(\frac{\dmhost}{\rm{ pc\,cm}^{-3}}\right) \approx 36\,\log{\left(\frac{\rm SFR}{M_{\odot}\,{\rm yr}^{-1}}\right)} + 84 
    \label{eq:dmh_sfr}
\end{equation}

\noindent with a Pearson correlation test giving a coefficient of $0.85$ (with $p$-value of 4 $\times$ $10^{-3}$).
%(taking into account the statistical uncertainties). 
A correlation with SFR could be expected given that galaxies with higher star formation activity should also have larger $S(H\alpha$), which is directly proportional to \dmhism\ in our estimations (Equations~\ref{eq:em}~and~\ref{eq:dmhism}). Indeed, despite the fact that we use local values of $S(H\alpha$) to estimate \dmhism, we have shown that for our sample both local and global values of $S(H\alpha$) are similar (see Figure~\ref{fig:local_vs_global}; see also Section~\ref{sec:systematic}). We also observe a positive correlation between \dmhhalo\ and SFR (Pearson coefficient $0.91$, with $p$-value of 1 $\times$ $10^{-3}$) which can again be explained by the existence of the SFMS for star-forming galaxies.

In order to investigate whether the correlations between \dmhism\ and stellar mass, and \dmhhalo\ and SFR are driven by the SFMS we have performed the following experiment. We used the nominal stellar mass values  of the galaxies in our sample (without considering uncertainties) and obtained values for their SFR assuming they lie right on top of the SFMS fit reported by \cite{sfrtomass} in their equation~$4$. From these, we obtained an expectation of the \halpha\ flux using equation~14 from \citet{SFRtohalpha}, from which we got an `expected' \dmhism\ given the new \sbha, and by following our methodology (Section~\ref{sec:dmism}) but for a global measurement rather than a local measurement (see also Section~\ref{sec:gvsl}). From such an experiment we find a correlation between \dmhism\
and \dmhhalo\ that is stronger than our actual case (using the actual local \halpha\ measurements) and also stronger when we compare it with our actual global measurements. This indicates that indeed, the SFMS can account for all the correlation observed between \dmhism\ and \dmhhalo, and hence both \dmhism\ and stellar mass, and \dmhhalo\ and SFR.

Finally, we also explore a possible correlation of \dmhdirect\ with the projected distance of the FRB sightline with respect to the galaxy halo center. In the right of Figure~\ref{fig:mass_sfr_offset} we show \dmhdirect\ as a function of the projected distance of the FRB sightline to the center of their host galaxy (black points), again also showing their \dmhhalo\ (dark blue points) and \dmhism\ (light blue points) contributions. In contrast to previous cases, we do not observe any strong correlation between \dmhdirect\ and FRB projected distances (we only report a weak anti-correlation) which can be explained by the fact that the most (least) massive galaxies have their FRBs farther (closer) to their centers. This result is driven by the adopted mNFW profiles which produce a rather flat (weak anti-correlation) between \dmhhalo\ and projected distance (see \citealp{ProchaskaZheng2019}). Furthermore, the fact that FRBs tend to lie well within the center of the dark matter halos of their host galaxies indicate that the \dmhhalo\ is indeed dominated by the halo mass (i.e. stellar mass) rather than projected distance. 
%\nt{can we conclude something else?}

Table~\ref{tab:params} in the Appendix summarizes the parameters of our fits and correlation coefficients discussed in this section.

\subsection{Possible redshift evolution of \dmhdirect\ and redshift biases}
\label{sec:redshift}

We also investigate a possible redshift dependence in our \dmhost\ estimates. Figure~\ref{fig:redshift} shows \dmhost\ as a function of redshift (top panel), where the black and red points correspond to \dmhdirect\ and \dmhmacq, respectively. We observe no strong redshift evolution in the individual values of \dmhost. The black line corresponds to a fit of the form,

\begin{equation}
    \dmhdirect(z) =\rm DM _{{\rm host},0} (1+z)^\alpha \ \ ,
    \label{eq:zfit}
\end{equation}

\noindent which shows a flat exponent ($\alpha \approx 0.3 \pm 1.7$) and a DM$_{{\rm host},0}\approx 73 \pm 33$\,\pccc, consistent with our $\langle \dmhdirect \rangle$. 
The apparent lack of redshift evolution indicates that \dmhost\ is driven by internal galaxy properties (e.g. stellar mass and SFR) and thus no extra corrections for redshift (galaxy evolution) are needed, at least for $z\lesssim 0.5$. We remind the reader that this \dmhost\ estimate is in the rest-frame, thus a factor of $(1+z)$ should still be applied when considering \dmhobs\ (see Equation~\ref{eq:dm_obs}).

In contrast, the \dmhmacq\ estimations show a much larger scatter and a possible evolution with redshift, which may indicate a possible redshift bias. The bottom panel of Figure~\ref{fig:redshift} shows the difference between \dmhdirect\ and \dmhmacq\ as a function of redshift. Here, we also observe a possible bias as a function of redshift for individual measurements where \dmhmacq\ tend to be larger than our \dmhdirect\ at lower redshifts ($z\lesssim 0.3$), and lower than \dmhdirect\ at higher redshifts ($z\gtrsim 0.3$). This discrepancy could be due to an underestimation of the Macquart relation at lower redshifts, making \dmhmacq\ systematically larger (see Equation~\ref{eq:host_max}). Another possibility is that our calculation of \dmhdirect\ is underestimated at lower redshifts, caused by an unknown systematic effect; we note that the expected systematic error of $\sim 30\%$ (see Section~\ref{sec:systematic}) is not enough to make this apparent redshift bias disappear. Given that our sample size is still small, there is also the possibility of this being driven by low number statistics (e.g. \rf{the} current trend is partially induced by the two FRBs with the largest \dmhmacq, which also correspond to the two largest \rf{values of} \dmfrb), motivating future similar studies with larger samples.

\subsection{Comparison with previous work}
\label{sec:previous}

Our empirical estimation is consistent with recent observational results from \citet{khrykin2024b} who reported $\langle{\rm DM}_{\rm host} \rangle=90^{+29}_{-19}$\,\pccc\ applied to a sample of 8 FRB sightlines and using a different methodology. However, we note that 6 out of 8 of these were also included in our analysis, and thus these two results are not fully independent. \rf{Our empirical estimation is  also consistent with that of \citet{shin+23} who reported a median \rf{value} of \dmhost\ $=84^{+69}_{-49}$ \pccc\ for CHIME FRBs. In contrast, our results are inconsistent with those from \citet{wang+24} who reported much larger values of $\langle{\rm DM}_{\rm host} \rangle$ = $560^{+37}_{-190}$ \pccc\ for their power law density model, and $\langle{\rm DM}_{\rm host} \rangle$ = $490^{+74}_{-129}$ \pccc\ for their SFR model, also for CHIME FRBs. Our estimates are also lower than those reported by \citet{bhardwaj} based on FRB scattering timescales, of $\langle{\rm DM}_{\rm host} \rangle \gtrsim 170$\,\pccc\ for a range of galaxy inclinations.}

% theoretical work
In terms of recent theoretical work, \cite{mo2023} analyzed cosmological hydrodynamical simulations using two different models for the location of FRBs within the simulated host galaxies: (i) following the star-formation, and (ii) following the stellar mass. They obtained different \dmhost\ values for these different models, reporting medians of 179 and 63 \pccc\ (rest-frame), respectively. Our observational result lies between these two values, and is consistent with both of them given the reported uncertainties and large scatter across models.  \cite{mo2023} also report a linear relation between their inferred \dmhost\ and the logarithm of the host stellar mass for both models following a similar trend as the one reported here, although again with much larger variations depending on the model used (see also \citealt{zhang2020, kovacs}). Finally they report a possible evolution of \dmhost\ (rest-frame) with redshift using a power-law fit as we do here (Equation~\ref{eq:zfit}), but with an $\alpha$ varying between 0.8 and 1.8 depending on the model (see also \citealt{zhang2020, kovacs, orr+24}), whereas we do not see any evolution from our current sample (see Section~\ref{sec:redshift}). Therefore, empirical \dmhost\ estimations have the potential to constrain sub-grid physics of galaxy evolution (e.g. \citealp{khrykin2024b}) and also provide important clues for the origin of FRBs themselves.

\subsection{Systematic uncertainties}
\label{sec:systematic}

In this work we model \dmhdirect\ as the sum of two contributions: \dmhism\ and \dmhhalo\ (see Section~\ref{sec:empiric}). Out of these two estimates, we consider that \dmhism\ is more uncertain than \dmhhalo\ because of the several assumptions regarding geometry, temperature, and other properties of the interstellar medium, and such systematic effects are not fully accounted for in our reported statistical uncertainties; for instance, recent studies of pulsars in the Large Magellanic Cloud (LMC), report variations of observed DMs of several tens of \pccc\  \citep[e.g.][see their table~1]{prayag24}. \rf{In a more extreme situation, if FRBs originate outside the disks of galaxies (e.g. globular clusters; \citealt{Kirsten+22}), then \dmhism$\sim0$\,\pccc, at least for those cases in which the FRB is foreground to the disk itself.}
Although the \dmhhalo\ inferences also depend on several systematic factors, including the assumed model of baryons around galaxies (Equation~\ref{eq:pz19}), ionization fraction ($f_{\rm hot}$) and the path length of the FRB along the galaxy host halo, the fact that we sample \dmhhalo\ from a PDF that does include (already large) uncertainties in both the stellar mass and the dark matter halo mass, makes their systematic effect less important relative to those associated with \dmhism. 

In order to estimate such a systematic effect we have repeated our individual \dmhdirect\ measurements using $10^5$ realizations of \dmhism\ uniformly sampled from a PDF whose mean value is between $[0.5, 1.5]$ times our fiducial values (e.g. those reported in Table~\ref{tab:localvsglobal}, sixth column), leaving \dmhhalo\ unchanged (see Section~\ref{sec:dmdirect} and Equation~\ref{eq:dmhost}). We consider that such a range in variation includes most of the systematic uncertainties in \dmhism\ (including geometrical factors that appear in Equation~\ref{eq:dmhism}). Figure~\ref{fig:mass_sfr_offset_v2} in the Appendix shows the effect of this experiment in our \dmhdirect\ measurement, where the shaded orange region represents the $1\sigma$ uncertainty (systematic) around our fiducial trends, corresponding to a $\sim 30\%$ (systematic) impact on \dmhdirect. 

In the context of measuring \dmhism, one can consider both a global or local measurement of \sbha\ (see Section ~\ref{sec:gvsl}). In our sample we have empirically found that both give quite similar results (see Figure~\ref{fig:local_vs_global}) given their statistical uncertainties. Such a result is partly driven by some of our galaxies not being fully resolved (7 out of the \nsamp\ have half-light radii comparable with the $0.4$\arcsec\ radius used for our local measurement). Similarly, we find that the larger differences between the two are coming from FRBs that have larger projected distances with respect to the disks traced by \halpha\ (as expected), but these cases are a minority (namely \rf{FRB20190611B} and FRB20210117A). In other words, most of the FRBs in our sample are coming from galaxy disks (e.g. \citealp{mannings21}). The systematic difference between the two is small and only implies a $\approx$\ddm\% bias in \dmhism, which is already contained in the $\sim 30\%$ systematic uncertainty budget reported above. Therefore, we conclude that estimating \dmhism\ from a global \halpha\ measurement provides similar level of accuracy than using a local estimate for partially resolved galaxies. This is not to say that both are the same, but just an observational limitation of our adopted technique. \rf{Finally, considering a more extreme (yet unlikely) situation where \dmhism$=0$\,\pccc, then only our results for \dmhhalo\ (e.g. see Table~\ref{tab:params}) should be used as reference for \dmhdirect\ instead.}

\begin{table*}[ht]
\centering
\begin{tabular}{ccccccc}
\hline
\hline
    FRB name & H$\alpha^{local}$ & H$\alpha^{global}$ & 
    S(H$\alpha)^{local}$ & S(H$\alpha)^{global}$ & 
    DM$_{host}^{ISM,local}$ & 
    DM$_{host}^{ISM,global}$ \\
& (10$^{-18}$ erg s$^{-1}$& (10$^{-18}$ erg s$^{-1}$ &  (10$^{-18}$ erg s$^{-1}$ &(10$^{-18}$ erg s$^{-1}$ & (pc cm$^{-3}$) &(pc cm$^{-3}$) \\ & cm$^{-2}$)  &cm$^{-2}$ )&cm$^{-2}$arcsec$^{-2}$)&cm$^{-2}$arcsec$^{-2}$)&\\ (1) & (2) & (3) & (4) & (5) & (6) & (7) \\ 
\hline
FRB20180924B &      21.6$\pm3.0$ &    104.1$\pm 13.5$ &       154.8$\pm21.3$ &       316.5$\pm 42.2$ &               40.2$\pm2.7$ &             57.3$\pm4.3$ \\
FRB20190102C &      20.7$\pm2.2$ &     59.3$\pm 16.3$ &       157.2$\pm16.9$ &       103.2$\pm 27.5$ &               41.4$\pm2.2$ &             33.7$\pm4.3$ \\
FRB20190608B &     106.8$\pm5.2$ &  5292.8$\pm 525.7$ &       325.3$\pm15.9$ &       350.6$\pm 34.8$ &               68.8$\pm1.7$ &             71.5$\pm3.4$ \\
FRB20190611B &     <8.8$^{\dagger}$&      23.0$\pm 5.3$ &      <111.3$^{\dagger}$&       245.8$\pm 56.3$ &             <32.7$^{\dagger}$&             48.6$\pm5.5$ \\
FRB20190711A &      34.8$\pm8.2$ &     16.1$\pm 16.6$ &      467.9$\pm110.3$ &      313.5$\pm 323.6$ &               60.6$\pm7.6$ &            51.3$\pm17.3$ \\
FRB20190714A &      70.0$\pm2.4$ &    524.4$\pm 68.0$ &       353.1$\pm12.1$ &       498.6$\pm 64.6$ &               64.8$\pm1.1$ &             77.0$\pm4.8$ \\
FRB20191001A &    309.0$\pm25.2$ &  1765.6$\pm 261.9$ &     1453.1$\pm118.4$ &     1173.4$\pm 174.1$ &              131.8$\pm5.2$ &               118.3$\pm8.5$ \\
FRB20200430A &      11.1$\pm1.9$ &    168.2$\pm 13.2$ &         38.9$\pm6.7$ &       197.2$\pm 15.5$ &               22.9$\pm1.9$ &                51.6$\pm2.2$ \\
FRB20200906A &      11.1$\pm1.3$ &   686.4$\pm 114.3$ &        97.0$\pm11.1$ &       436.7$\pm 72.7$ &               30.7$\pm1.7$ &                64.9$\pm5.1$ \\
FRB20210117A &     <7.1$^{\dagger}$&      20.1$\pm 4.7$ &       <32.9$^{\dagger}$&         27.6$\pm 6.4$ &             <20.1$^{\dagger}$&             18.5$\pm1.9$ \\
FRB20210320C &      78.6$\pm3.7$ &    805.0$\pm 77.0$ &       561.4$\pm26.2$ &       602.4$\pm 57.6$ &               79.0$\pm1.8$ &                81.7$\pm4.0$ \\
FRB20210410D &       7.6$\pm2.1$ &      53.9$\pm 8.1$ &         35.1$\pm9.6$ &       134.7$\pm 20.3$ &               22.1$\pm2.8$ &                43.4$\pm2.9$ \\
\hline
\end{tabular}
\caption{Road to \dmhism. (1) Name of the FRB; (2) Local H$\alpha$ flux from MUSE cubes (see Section~\ref{sec:dmism}); (3) Global H$\alpha$ flux from MUSE cubes; (4) Local H$\alpha$ surface brightness, corrected for dust extinction and brightness dimming; (5) Global H$\alpha$ surface brightness; (6) Local ISM Dispersion Measure;(7) Global ISM Dispersion Measure.\\
\rf{$\dagger$: Non-detections are reported as $2\sigma$ upper limits.}} \label{tab:localvsglobal}
\end{table*}

\begin{table*}[ht]
\centering
\begin{tabular}{ccccccc}
\hline
\hline
    FRB name &   log(M$_{\star}$ / M$_{\odot}$) &      log(M$_{halo}$ / M$_{\odot}$) & \dmhhalo  & \dmhdirect  & \dmhmax &  \dmhmacq$\dagger$ \\
&& &  (pc cm$^{-3}$) &(pc cm$^{-3}$) & (pc cm$^{-3}$) &(pc cm$^{-3}$)\\(1) & (2) & (3) & (4) & (5) & (6) & (7)\\
\hline
FRB20180924B & 10.39$\pm0.02$ &  11.96$\pm0.1$ &         36$\pm3$ &            76$\pm4$ &              167 &                   -2 \\
FRB20190102C &  9.69$\pm0.09$ & 11.47$\pm0.14$ &         24$\pm2$ &            65$\pm3$ &              169 &                   16 \\
FRB20190608B & 10.56$\pm0.02$ & 12.09$\pm0.11$ &         32$\pm7$ &           101$\pm7$ &              241 &                  181 \\
FRB20190611B &  9.57$\pm0.12$ & 11.43$\pm0.15$ &         26$\pm3$ &            58$\pm4$ &               67 &                  -136 \\
FRB20190711A &  9.10$\pm0.15$ &  11.23$\pm0.2$ &         25$\pm4$ &            86$\pm8$ &              332 &                   34 \\
FRB20190714A & 10.22$\pm0.04$ &  11.79$\pm0.1$ &         28$\pm2$ &            93$\pm2$ &              394 &                  273 \\
FRB20191001A & 10.73$\pm0.07$ & 12.37$\pm0.21$ &        44$\pm19$ &          176$\pm20$ &              391 &                  270 \\
FRB20200430A &  9.30$\pm0.07$ & 11.22$\pm0.14$ &         17$\pm2$ &            40$\pm2$ &              285 &                  203 \\
FRB20200906A & 10.37$\pm0.05$ & 11.96$\pm0.11$ &         38$\pm4$ &            69$\pm4$ &              438 &                  240 \\
FRB20210117A &  8.59$\pm0.05$ & 10.91$\pm0.15$ &         14$\pm1$ &            34$\pm2$ &              678 &                  569 \\
FRB20210320C & 10.37$\pm0.05$ & 11.93$\pm0.11$ &         34$\pm3$ &           113$\pm4$ &              224 &                   78 \\
FRB20210410D &  9.47$\pm0.05$ & 11.29$\pm0.13$ &         17$\pm1$ &            39$\pm3$ &              481 &                  409 \\
\hline
\end{tabular}
\caption{Summary of our main results (including road to \dmhhalo). (1) Name of the FRB; (2) Stellar mass of host galaxy; (3) Halo mass of host galaxy (see Section~\ref{sec:dmhhalo}); (4) Dispersion measure of the halo of the host galaxy (see Section~\ref{sec:dmhhalo}); (5) Empirical dispersion measure of the host galaxy (our main result; see Section~\ref{sec:dmdirect}); (6) Upper limit for the dispersion measure of the host galaxy (see Section~\ref{sec:macquart}); (7) Average dispersion measure of the host galaxy estimated from the Macquart relation (see Section~\ref{sec:macquart}).\\
$\dagger$\rf{:} Although negative values are nonphysical, we nevertheless report them as such for statistical consistency when taking the average of the ensemble.}
\label{tab:summary}
\end{table*}

\section{Summary and Conclusions}
\label{sec:summary}

In this work, we have empirically estimated the dispersion measure of FRB host galaxies, \dmhost, using a sample of \nsamp\ well-localized FRBs at redshifts $0.11<z<0.53$ observed with VLT/MUSE. For this estimation, we employ a method that considers the stellar mass of the host galaxies and the projected distance to the FRBs in order to infer the contribution from the host halos, \dmhhalo, and the \halpha\ nebular emission line flux to infer the contribution from their interstellar medium (ISM), \dmhism. The sum of these two contributions is our direct estimation of \dmhost, referred to as \dmhdirect\ (see Equation~\ref{eq:dmhost}). For comparison purposes we also estimate the inferred \dmhost, assuming that the sightlines to \rm{the} FRBs have an intergalactic medium DM contribution given by the Macquart relation, referred to as \dmhmacq\ (see Equation~\ref{eq:host}). We summarize our main results as follows:

\begin{enumerate}
\item We find that on average $\langle$DM$_{\rm host} \rangle = \avgdirect$\,\pccc\ \rf{with a standard deviation of \stddirect\,\pccc} (the uncertainty here is only statistical). We have estimated a systematic uncertainty of $\sim 30\%$ in our results, driven mostly by the unknown physical properties of the host ISM gas through which the FRB pulse propagated (including the immediate environment around the progenitor). Our reported \dmhdirect\ is larger than the typically used (fixed) value of $50$\,\pccc\ but consistent within uncertainties.\smallskip

\item We find no strong correlation between \dmhdirect\ and \dmhmacq, indicating that our current modeling could be missing an important DM contribution to the \dmfrb\ for a fraction of FRBs (e.g. intrinsic variations to \dmhost\ and/or unaccounted intervening halos or cosmic web structures). \smallskip

\item We find a positive correlation between \dmhdirect\ with stellar mass and SFR of the host galaxy (see Figure~\ref{fig:mass_sfr_offset}) of the form given by Equations~\ref{eq:dmh_mass}~and~\ref{eq:dmh_sfr} (parameters summarized in Table~\ref{tab:params}). In contrast, no strong correlation is observed between \dmhdirect\ and FRB projected distance with respect to the host centers (see Section~\ref{sec:correlations}).\smallskip

\item We do not find any strong redshift evolution of \dmhost\ (rest-frame) in our sample (Figure~\ref{fig:redshift} and Section~\ref{sec:redshift}). However, we find a possible redshift bias associated with the indirect estimation of \dmhost\ based on the Macquart relation, and/or produced by an unknown bias in our \dmhdirect\ measurements.

\end{enumerate}

Our reported correlations and the lack of a strong redshift evolution can be used to constrain models for the progenitor of FRBs, e.g. by comparing them with theoretical models. However, larger samples are needed to keep improving on the present results and thus provide more precise and accurate observed values of \dmhost.

%----------------------------------------------------------------

\begin{acknowledgements}

\rf{We thank the anonymous referee for their careful revision and suggestions that improved the paper.} We thank Ron Ekers and Kasper Heintz for useful comments and discussions. 
This work is based on observations collected at the European Southern Observatory under ESO programmes 2102.A-5005, 0104.A-0411, 0105.20HG and 0110.241Y.
LBC and NT acknowledge support by FONDECYT grant 11191217.
LBC acknowledge support by Beca Postgrado PUCV 2022-2023.
ISK and NT acknowledge the support received by the Joint Committee ESO-Government of Chile grant ORP 40/2022.
LBC, NT, ISK, and JXP, as members of the Fast and Fortunate for FRB Follow-up team, acknowledge support from NSF grants AST-1911140, AST-1910471 and AST-2206490. LM is supported by an Australian Government Research Training Program (RTP) Scholarship. RMS acknowledges support through Australian Research Council Future Fellowship FT190100155 and Discovery Project DP220102305.

\end{acknowledgements}

\appendix
\section{\halpha\ emission for the rest of the sample}

\begin{figure}[h]
\centering
\includegraphics[width=0.45\textwidth]{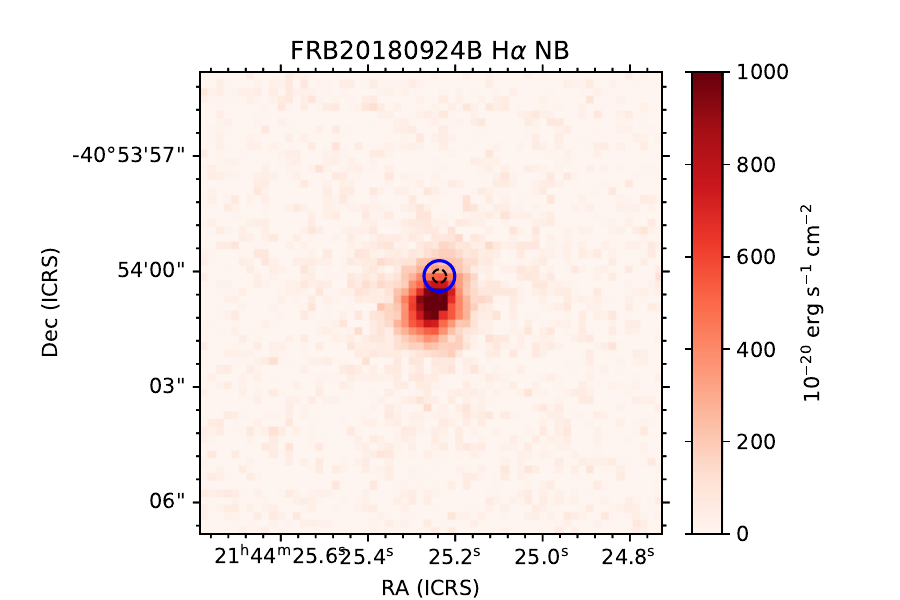}
\includegraphics[width=0.45\textwidth]{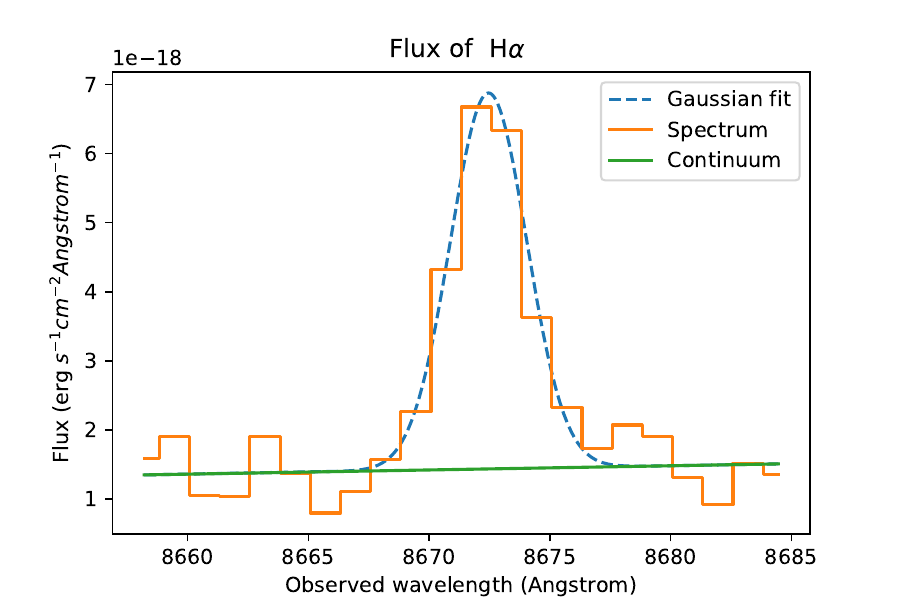}
\caption{Same as Figure~\ref{fig:halpha} but for FRB20180924B.}
 \label{fig:halpha1}
\end{figure}

\begin{figure}[h]
\centering
\includegraphics[width=0.45\textwidth]{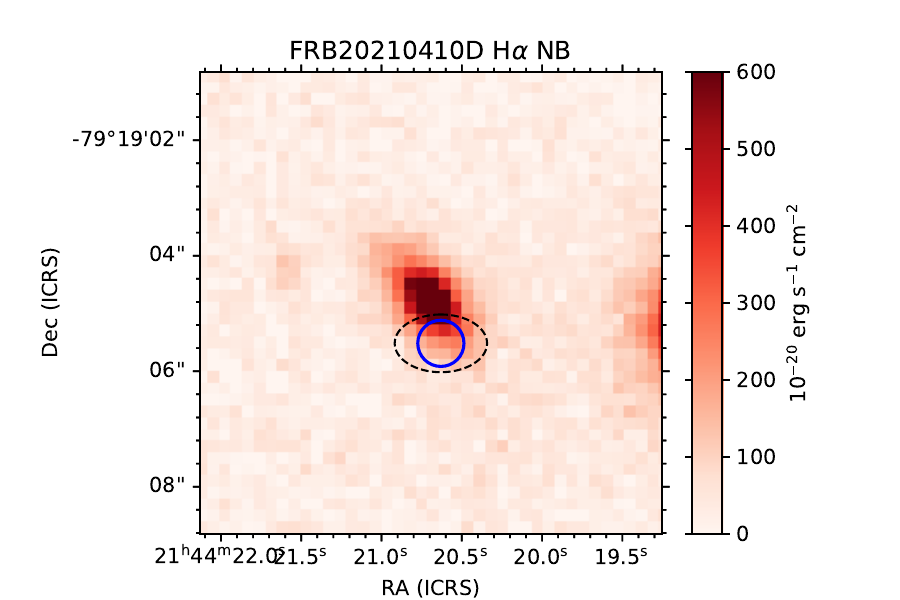}
\includegraphics[width=0.45\textwidth]{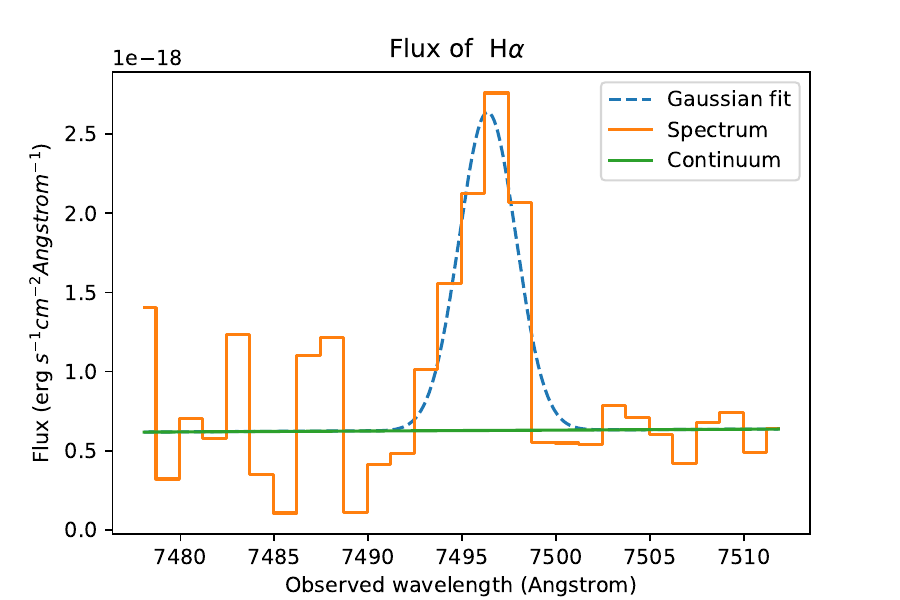}
\caption{Same as Figure~\ref{fig:halpha} but for FRB20210410D.}
 \label{fig:halpha12}
\end{figure}

In Figures \ref{fig:halpha1} to \ref{fig:halpha12} we show, as in Figure \ref{fig:halpha} the location of the FRB within the host galaxy, along with the corresponding extracted spectrum for the rest of the sample, excluding FRB20191001A shown already in Figure~\ref{fig:halpha}. The galaxy map corresponds to the measured emission line (\halpha, H$\beta$, H$\gamma$, as appropriate).

\begin{figure}[h]
\centering
\includegraphics[width=0.45\textwidth]{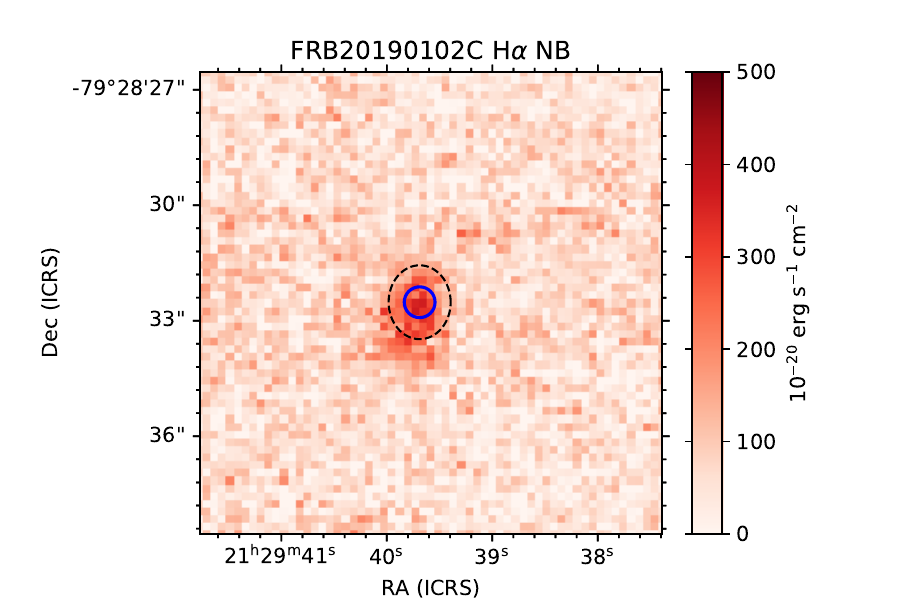}
\includegraphics[width=0.45\textwidth]{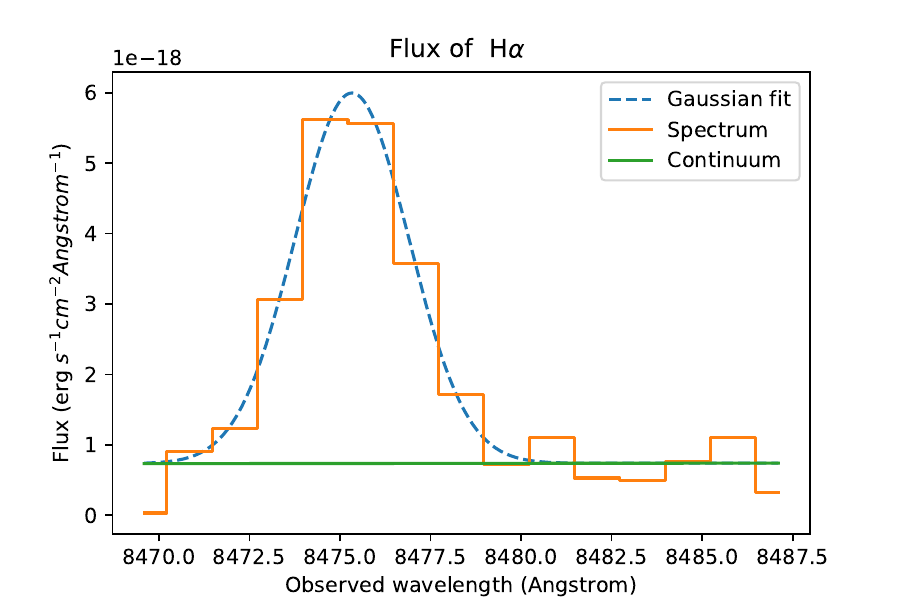}
\caption{Same as Figure~\ref{fig:halpha} but for FRB20190102C.}
\end{figure}

\begin{figure}[h]
\centering
\includegraphics[width=0.45\textwidth]{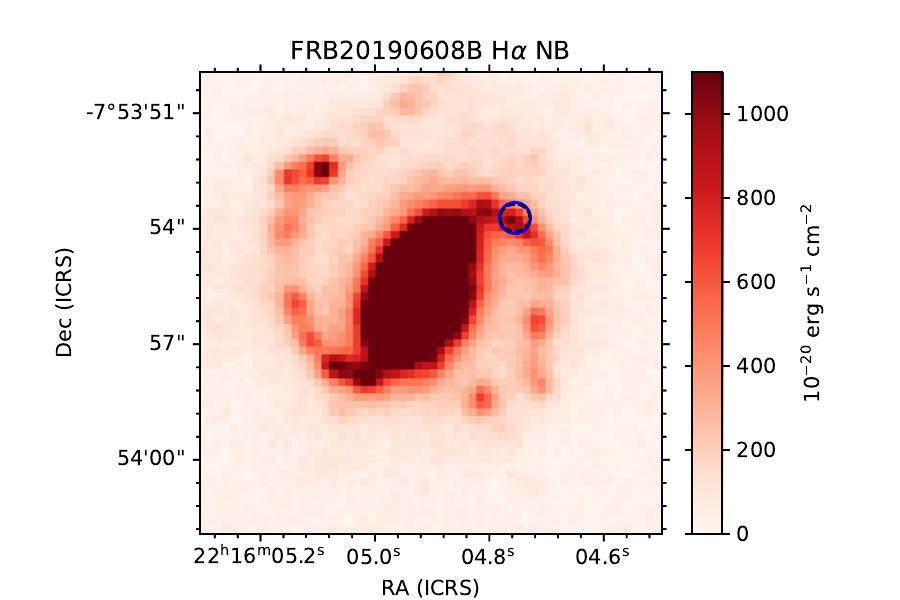}
\includegraphics[width=0.45\textwidth]{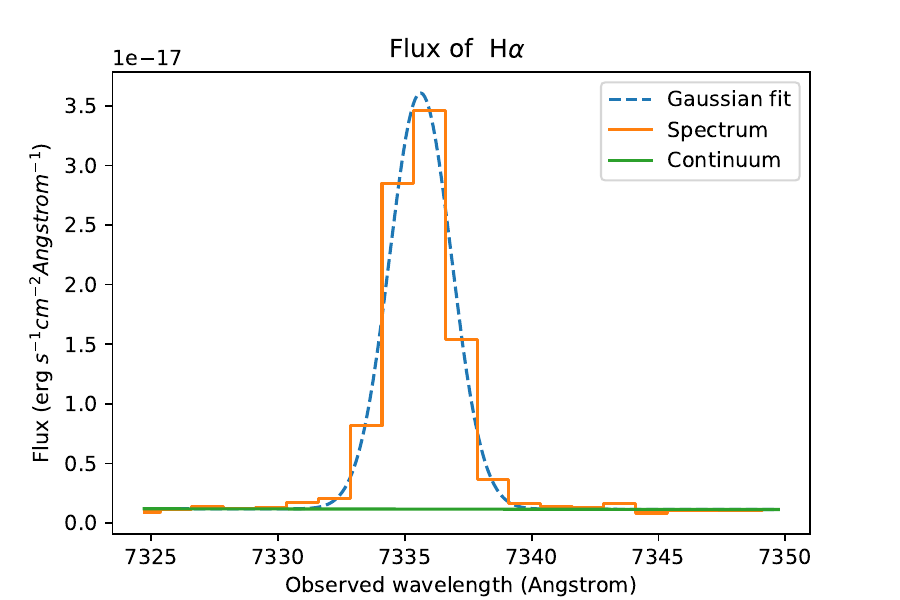}
\caption{Same as Figure~\ref{fig:halpha} but for FRB20190608B.}
\end{figure}

\begin{figure}[h]
\centering
\includegraphics[width=0.45\textwidth]{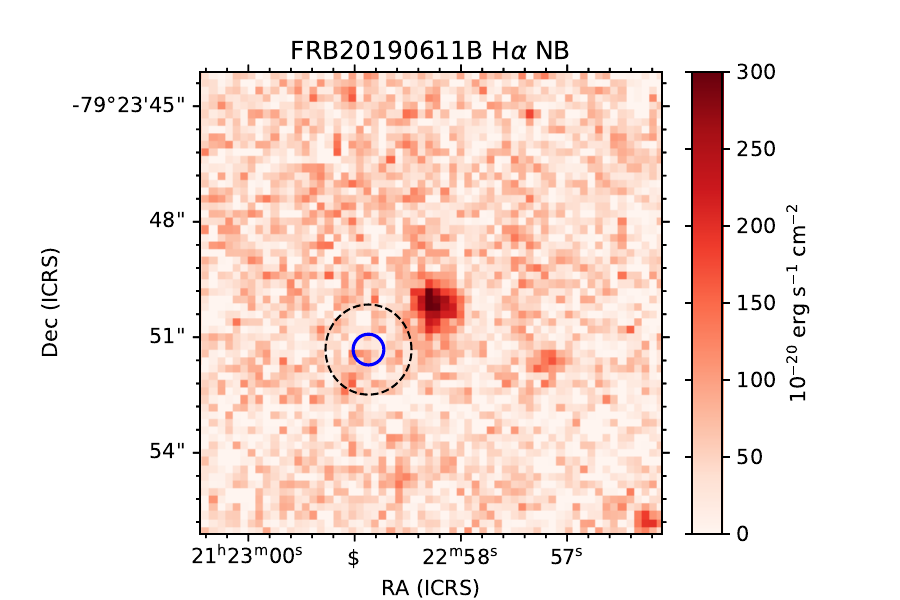}
\includegraphics[width=0.45\textwidth]{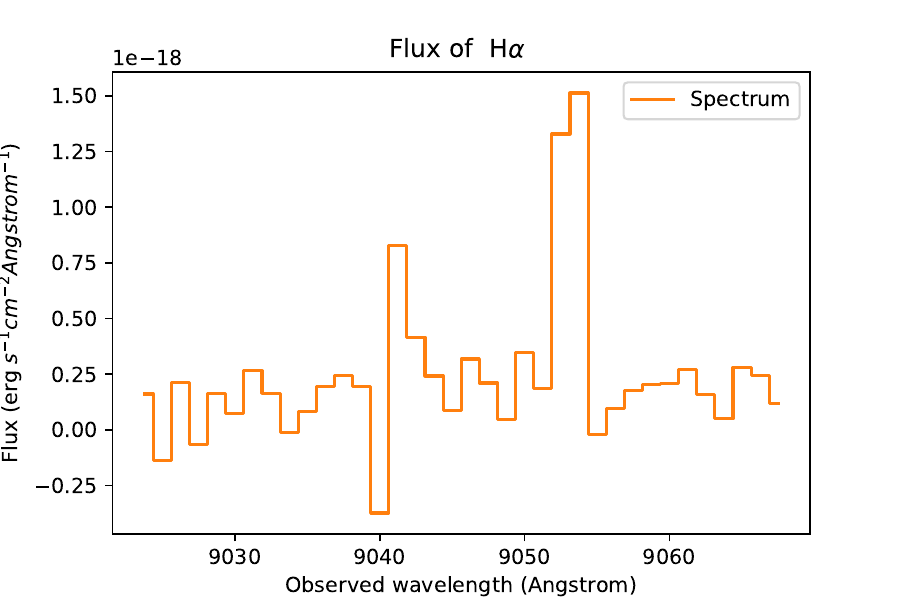}
\caption{Same as Figure~\ref{fig:halpha} but for FRB20190611B. \rf{In this case, we cannot fit a Gaussian and thus just report a $2\sigma$ upper limit.}}
\end{figure}

\begin{figure}[h]
\centering
\includegraphics[width=0.45\textwidth]{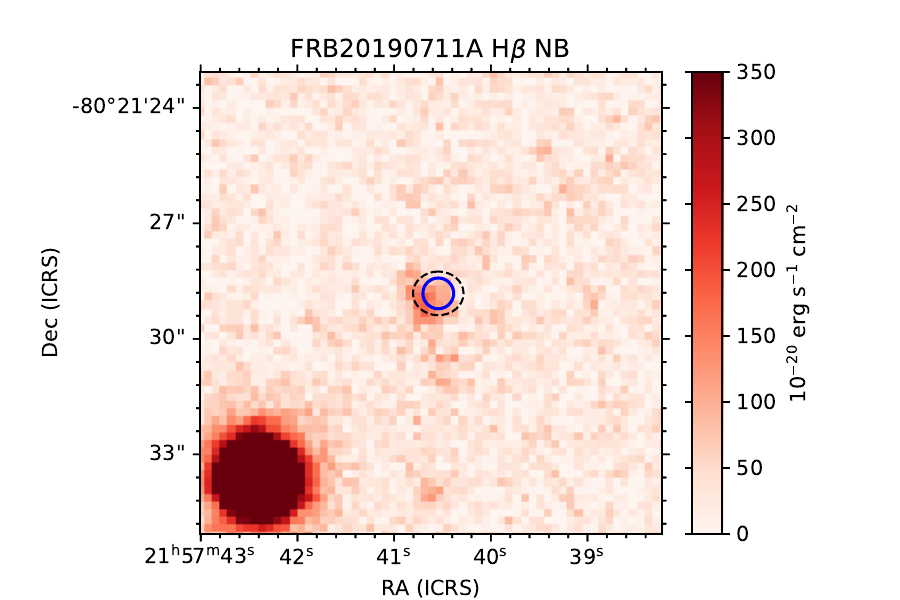}
\includegraphics[width=0.45\textwidth]{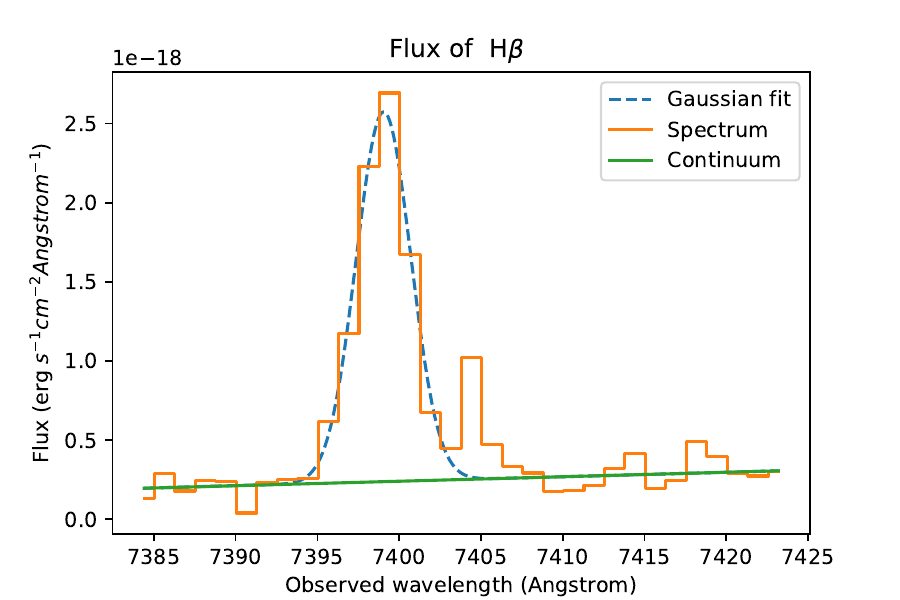}
\includegraphics[width=0.45\textwidth]{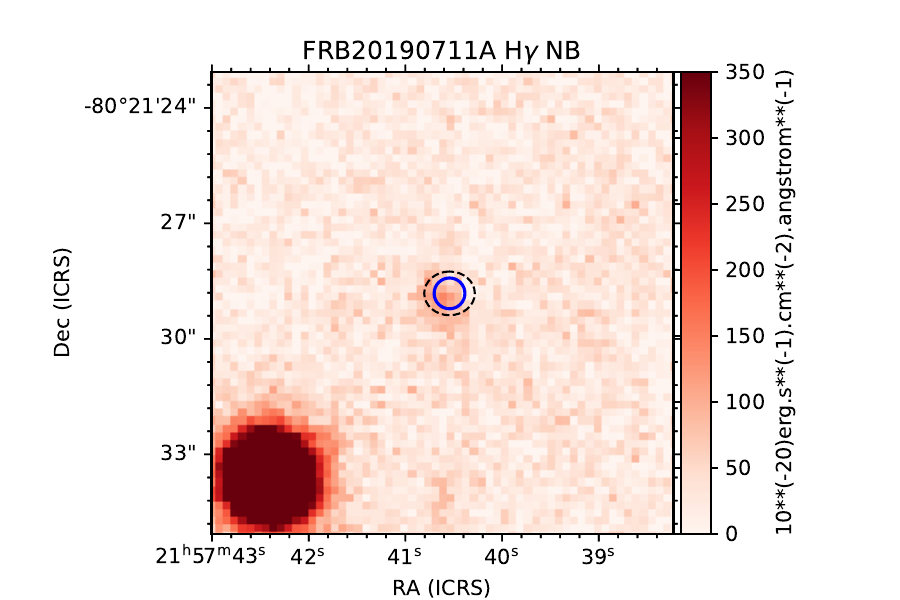}
\includegraphics[width=0.45\textwidth]{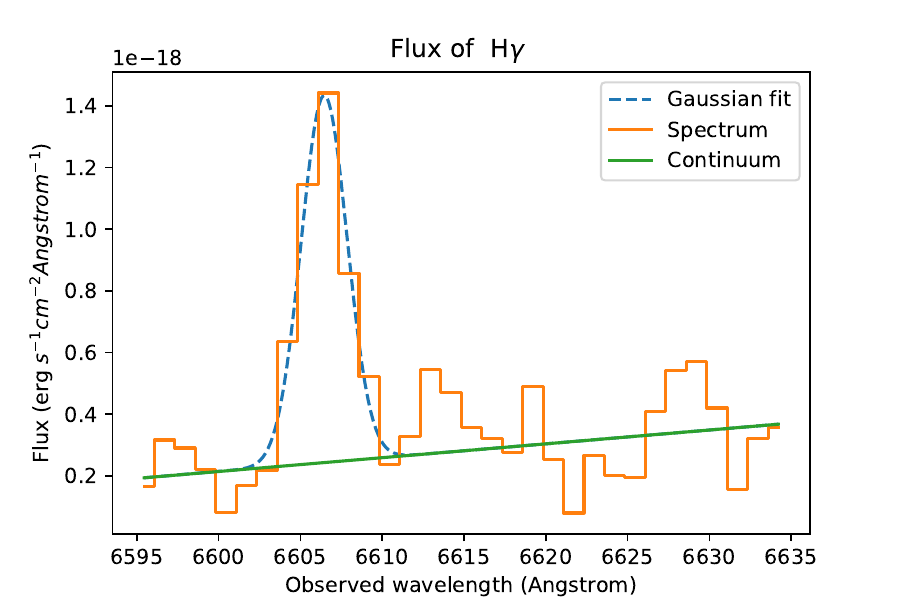}
\caption{Same as Figure~\ref{fig:halpha} but for FRB20190711A. In this case, instead of using H$\alpha$ we use H$\beta$ and H$\gamma$.}
\end{figure}

\begin{figure}[h]
\centering
\includegraphics[width=0.45\textwidth]{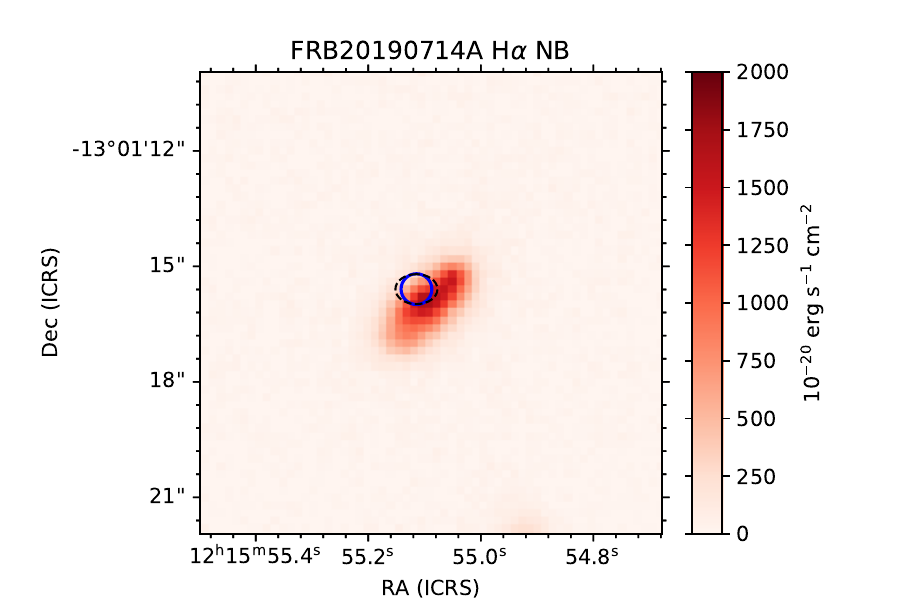}
\includegraphics[width=0.45\textwidth]{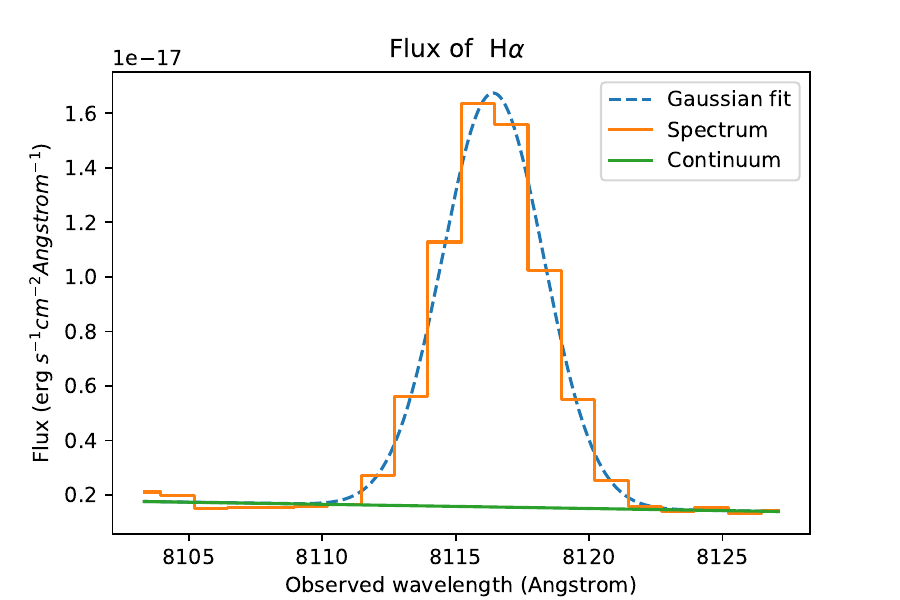}
\caption{Same as Figure~\ref{fig:halpha} but for FRB20190714A.}
\end{figure}

%\begin{figure}[h]
%\centering
%\includegraphics[width=0.4\textwidth]{ha_locales/FRB191001Halpha_location.pdf}
%\includegraphics[width=0.4\textwidth]{ha_locales/FRB20191001AHalpha.pdf}
%\caption{Same as Figure~\ref{fig:halpha} but for FRB20191001A.}
%\end{figure}

\begin{figure}[h]
\centering
\includegraphics[width=0.45\textwidth]{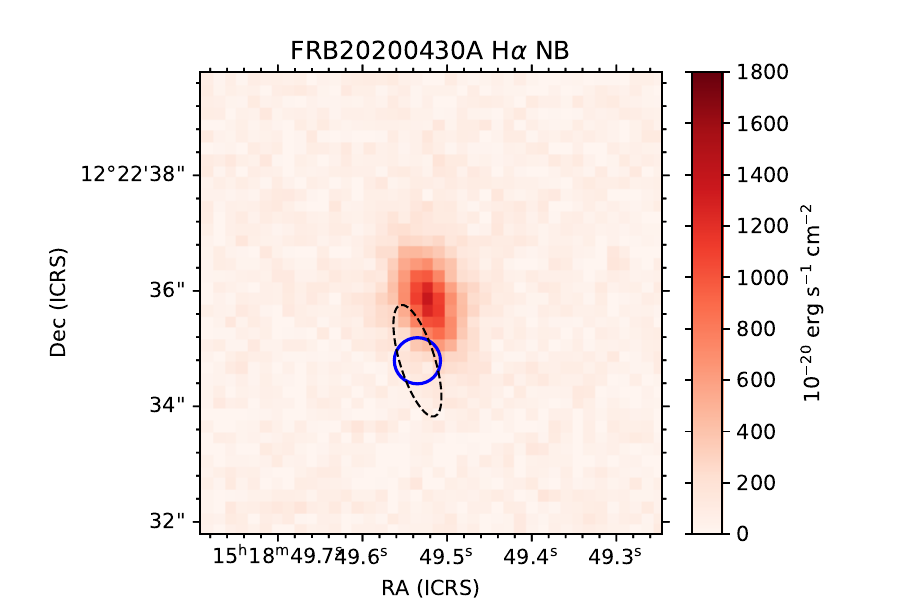}
\includegraphics[width=0.45\textwidth]{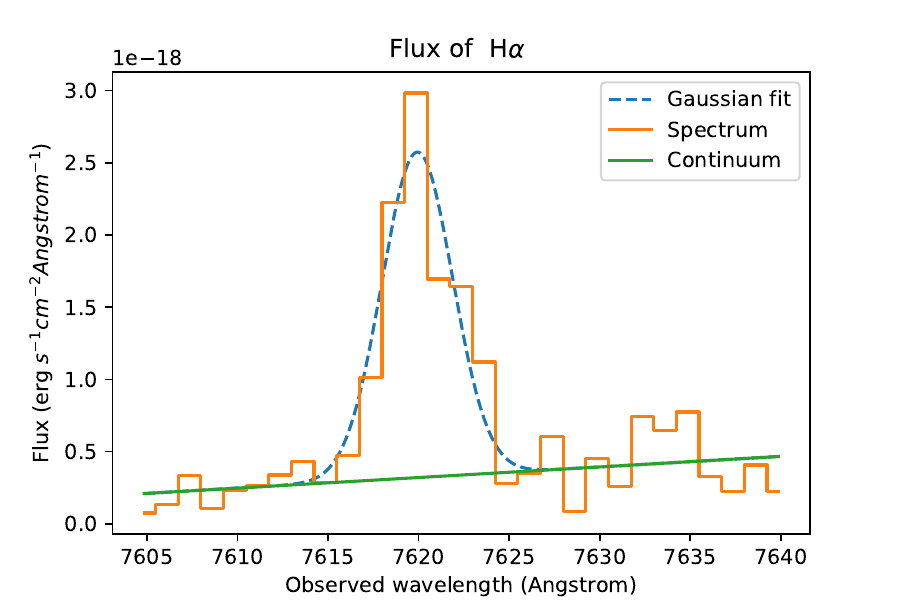}
\caption{Same as Figure~\ref{fig:halpha} but for FRB20200430A.}
\end{figure}

\begin{figure}[h]
\centering
\includegraphics[width=0.45\textwidth]{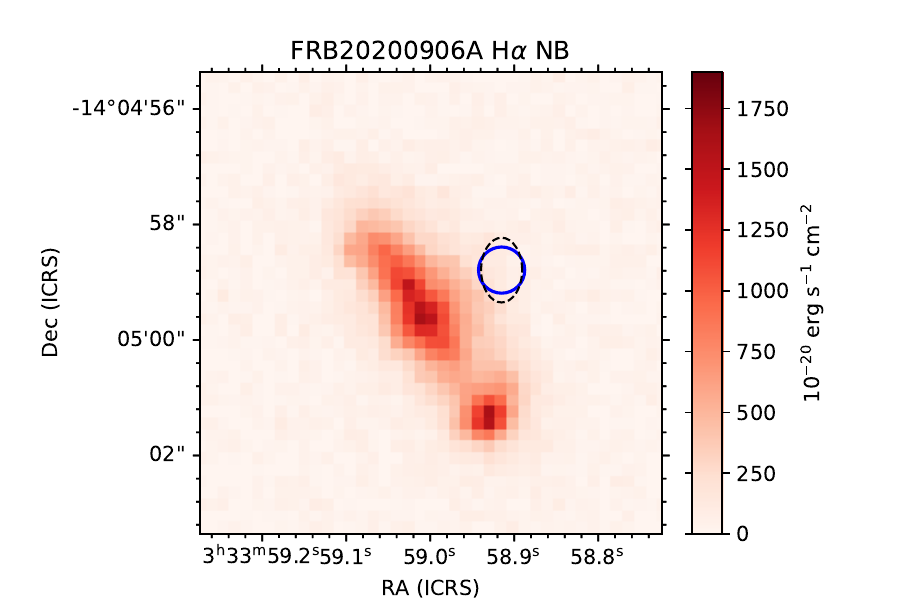}
\includegraphics[width=0.45\textwidth]{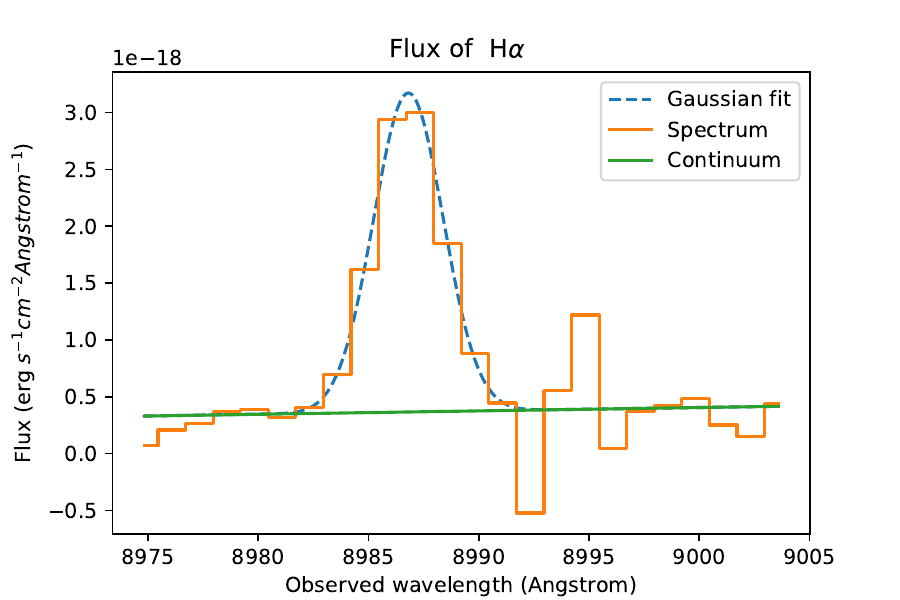}
\caption{Same as Figure~\ref{fig:halpha} but for FRB20200906A.}
\end{figure}

\begin{figure}[h]
\centering
\includegraphics[width=0.45\textwidth]{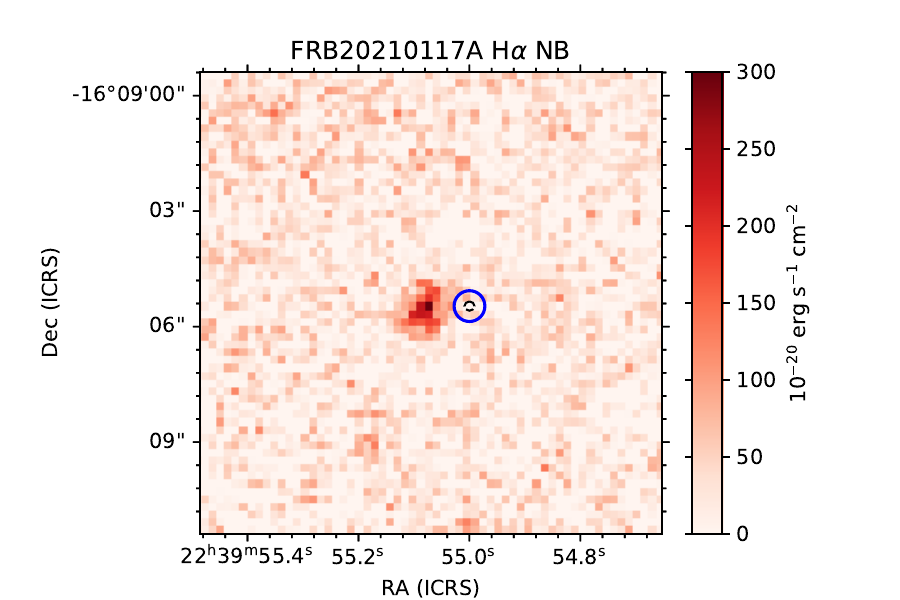}
\includegraphics[width=0.45\textwidth]{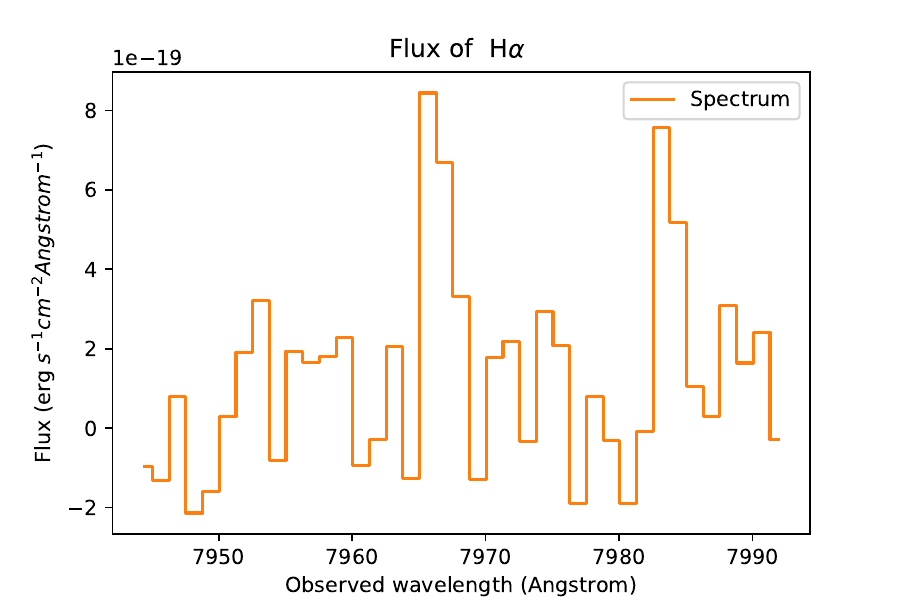}
\caption{Same as Figure~\ref{fig:halpha} but for FRB20210117A. \rf{In this case, we cannot fit a Gaussian and thus just report a $2\sigma$ upper limit.}}
\end{figure}

\begin{figure}[h]
\centering
\includegraphics[width=0.45\textwidth]{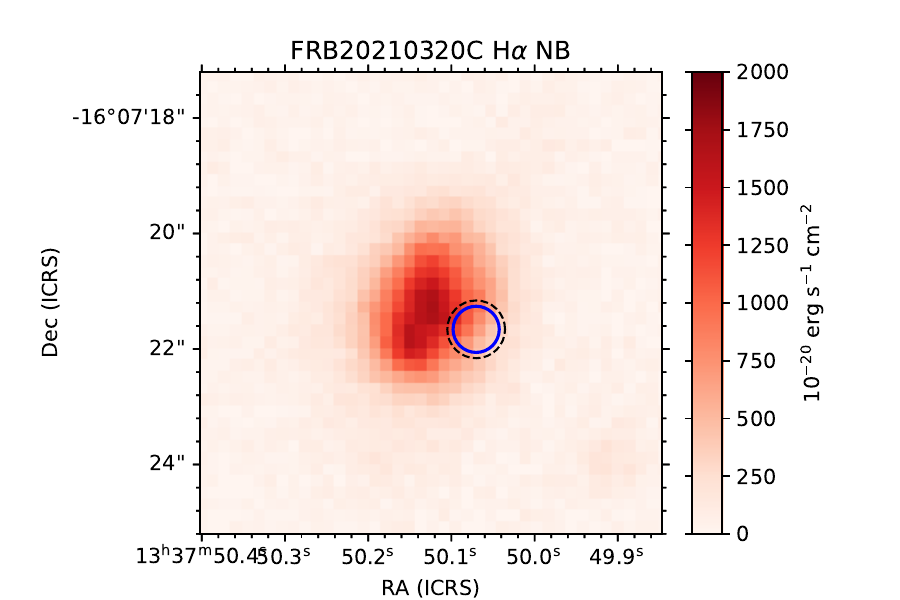}
\includegraphics[width=0.45\textwidth]{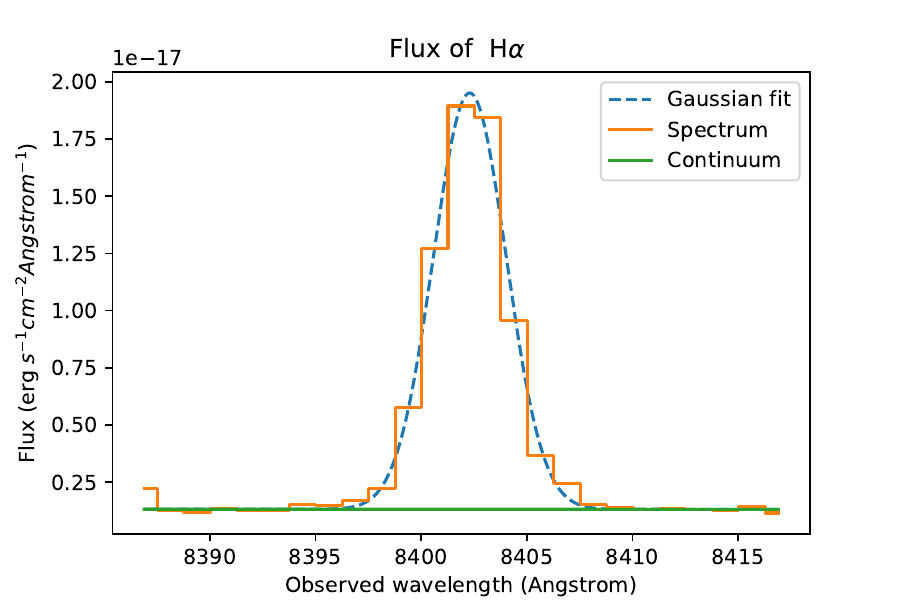}
\caption{Same as Figure~\ref{fig:halpha} but for FRB20210320C.}
\end{figure}

%------------------------------------------------------------

\section{Parameters of the reported fits}
In Table~\ref{tab:params} we present a summary of the parameters reported in our fits of Section~\ref{sec:correlations} .

\begin{table*}[ht]
\centering
\begin{tabular}{c|cccc|cccc|cccc}
\hline
\hline
Component& \multicolumn{4}{ c| }{M$_{\star}$} &\multicolumn{4}{ c| }{SFR } &\multicolumn{4}{ c }{Offset} \\
$y$ &   $b$ &   $m$ &Pearson &$p$ & $b$ &   $m$ &Pearson &$p$ &  $b$ &  $m$& Pearson&$p$\\
& (\pccc) &&coeff.&value & (\pccc) &&coeff. &value&  (\pccc)&&coeff.& value\\
(1)&\multicolumn{4}{ c| }{(2)}&\multicolumn{4}{ c| }{(3)}&\multicolumn{4}{ c }{(4)}\\
\hline
\dmhdirect  &            86$\pm 8$ &          43$\pm 13$ & 0.73 &7$\times$$10^{-2}$&     84$\pm 5$ &         36$\pm 7$ & 0.85 &4$\times$$10^{-3}$&       86$\pm 21$ &        -2$\pm 4$ & 0.22&0.71 \\
\dmhhalo    &            30$\pm 1$ &          12$\pm 2$  & 0.89 &1$\times$$10^{-3}$&     30$\pm 1$ &          9$\pm 1$ & 0.91 &1$\times$$10^{-3}$&       26$\pm  5$ &         0$\pm 1$ & 0.13&0.69 \\
\dmhism     &            55$\pm 8$ &         31$\pm 12$  & 0.64 &3$\times$$10^{-2}$&     55$\pm 6$ &         27$\pm 7$ & 0.78 &3$\times$$10^{-3}$&       60$\pm 14$ &        -2$\pm 3$ &-0.18&0.57 \\
\hline
\end{tabular}
\caption{Parameters of linear fits  of the form $y = m x + b$ for correlations shown in Figure~\ref{fig:mass_sfr_offset} (see also Section~\ref{sec:correlations}) \rf{and their corresponding Pearson coefficients and $p$-values.}. (1) Component fitted; (2) Parameters for stellar mass fits, with $x= {\rm log(M_{\star} /10^{10}M}_{\odot})$; (3) Parameters for SFR fits, with $x$ = {\rm log(SFR /M}$_{\odot}${\rm yr}$^{-1})$; (4) Parameters for projected distance fits, with $ x $ being the impact parameter between the FRB and the galaxy center in kpc
. }
\label{tab:params}
\end{table*}

\section{Systematic uncertainty estimation of our reported trends}
\begin{figure*}
    \includegraphics[width=0.33\textwidth]{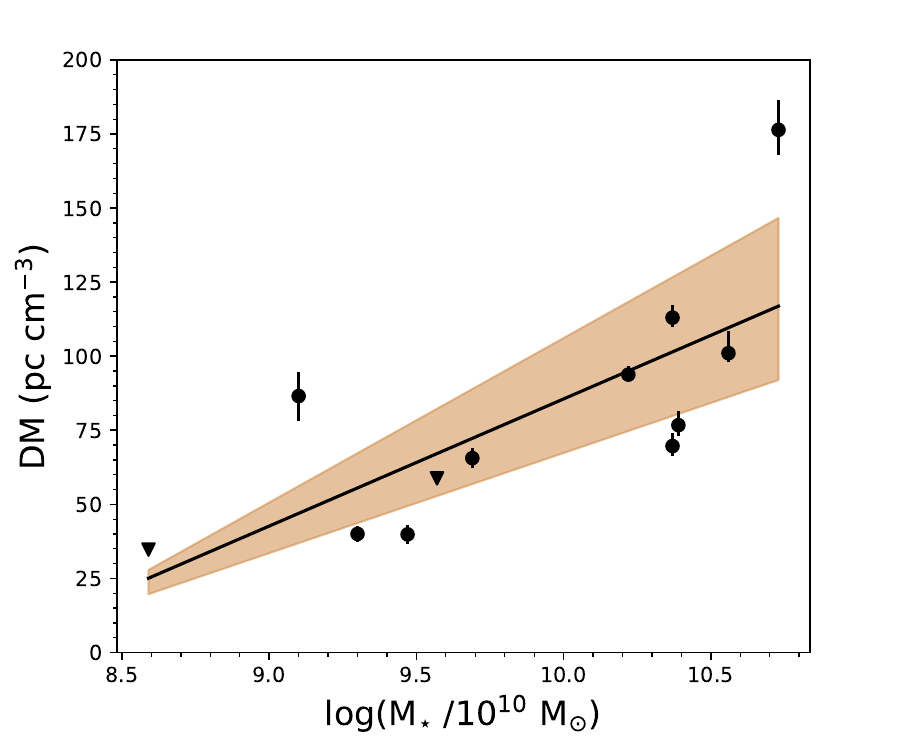}
    \includegraphics[width=0.33\textwidth]{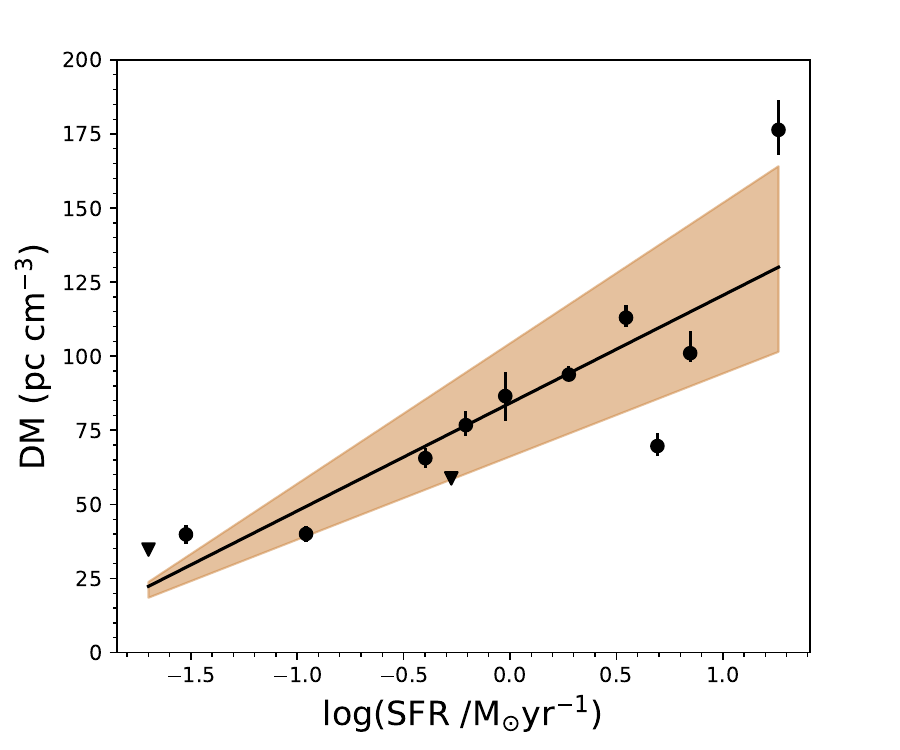}
    \includegraphics[width=0.33\textwidth]{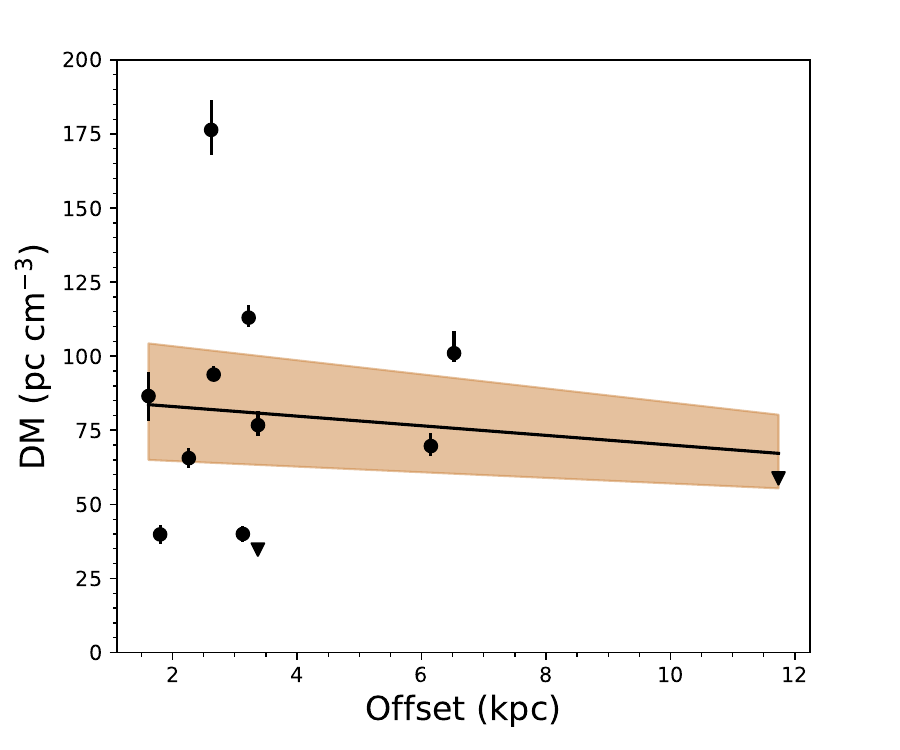}
    \caption{\dmhdirect\ vs. stellar mass (in log scale; left), SFR (in log scale; middle) and projected distance (right) of the host galaxies. The black points and the black lines are the same as those in Figure~\ref{fig:mass_sfr_offset}. The shaded area corresponds to our estimated systematic uncertainty (see Section~\ref{sec:systematic}).
    } \label{fig:mass_sfr_offset_v2}
\end{figure*}

In Figure~\ref{fig:mass_sfr_offset_v2} we shown our estimated systematic uncertainties (shaded brown regions) for the different reported trends of Section~\ref{sec:correlations}. Details on how this is done are given in in Section~\ref{sec:systematic}.

\bibliography{references}
\bibliographystyle{aa}

\end{document}